\documentclass[sigconf,noacm]{acmart}

\usepackage[ruled, vlined, linesnumbered]{algorithm2e}
\usepackage{algorithmic}
\usepackage{latexsym}
\usepackage{amsfonts}
\usepackage{comment}
\usepackage{caption}
\usepackage{subfigure}
\usepackage{multirow}
\usepackage{subfigmat}

\makeatletter
\newcommand{\figcaption}[1]{\def\@captype{figure}\caption{#1}}
\newcommand{\tblcaption}[1]{\def\@captype{table}\caption{#1}}
\makeatother

\AtBeginDocument{%
  }

\copyrightyear{2023}
\acmYear{2023}
\setcopyright{rightsretained}
\acmConference[aiDM '23]{Sixth International Workshop on Exploiting Artificial Intelligence Techniques for Data Management }{June 18, 2023}{Seattle, WA, USA}
\acmBooktitle{Sixth International Workshop on Exploiting Artificial Intelligence Techniques for Data Management (aiDM '23), June 18, 2023, Seattle, WA, USA}
\acmDOI{10.1145/3593078.3593932}
\acmISBN{979-8-4007-0193-1/23/06}

\begin{document}

%%
%% The "title" command has an optional parameter,
%% allowing the author to define a "short title" to be used in page headers.
\title{Learned Spatial Data Partitioning}

%%
%% The "author" command and its associated commands are used to define
%% the authors and their affiliations.
%% Of note is the shared affiliation of the first two authors, and the
%% "authornote" and "authornotemark" commands
%% used to denote shared contribution to the research.
\author{Keizo Hori}
\email{hori.keizo@ist.osaka-u.ac.jp}
\affiliation{%
  \institution{Osaka University}
  \city{Suita}
  \state{Osaka}
  \country{Japan}
}
\author{Yuya Sasaki}
\email{sasaki@ist.osaka-u.ac.jp}
\affiliation{%
  \institution{Osaka University}
  \city{Suita}
  \state{Osaka}
  \country{Japan}
}
\author{Daichi Amagata}
\email{amagata.daichi@ist.osaka-u.ac.jp}
\affiliation{%
  \institution{Osaka University}
  \city{Suita}
  \state{Osaka}
  \country{Japan}
}
\author{Yuki Murosaki}
\email{murosaki.yuki@ist.osaka-u.ac.jp}
\affiliation{%
  \institution{Osaka University}
  \city{Suita}
  \state{Osaka}
  \country{Japan}
}
\author{Makoto Onizuka}
\email{onizuka@ist.osaka-u.ac.jp}
\affiliation{%
  \institution{Osaka University}
  \city{Suita}
  \state{Osaka}
  \country{Japan}
}

%%
%% By default, the full list of authors will be used in the page
%% headers. Often, this list is too long, and will overlap
%% other information printed in the page headers. This command allows
%% the author to define a more concise list
%% of authors' names for this purpose.
\renewcommand{\shortauthors}{Trovato et al.}

%%
%% The abstract is a short summary of the work to be presented in the
%% article.
\begin{abstract}
    Due to the significant increase in the size of spatial data, it is essential to use distributed parallel processing systems to efficiently analyze spatial data.
In this paper, we first study learned spatial data partitioning, which effectively assigns groups of big spatial data to computers based on locations of data by using machine learning techniques.
We formalize spatial data partitioning in the context of reinforcement learning and develop a novel deep reinforcement learning algorithm.
Our learning algorithm leverages features of spatial data partitioning and prunes ineffective learning processes to find optimal partitions efficiently.
Our experimental study, which uses Apache Sedona and real-world spatial data, demonstrates that our method efficiently finds partitions for accelerating distance join queries and reduces the workload run time by up to 59.4\%.

\end{abstract}

%%
%% The code below is generated by the tool at http://dl.acm.org/ccs.cfm.
%% Please copy and paste the code instead of the example below.
%%


%%
%% Keywords. The author(s) should pick words that accurately describe
%% the work being presented. Separate the keywords with commas.
%%\keywords{datasets, neural networks, gaze detection, text tagging}
%% A "teaser" image appears between the author and affiliation
%% information and the body of the document, and typically spans the
%% page.

%%
%% This command processes the author and affiliation and title
%% information and builds the first part of the formatted document.
\maketitle

\section{Introduction}
\label{introduction}
A large amount of spatial data, such as temperature data, traffic logs, and geo-tagged blog posts, is drastically increasing.
%The recent development of communication networks provides mobile and sensor devices to connect the Internet from anywhere. 
%Mobile and sensor devices generate a large amount of spatial data, such as temperature data, traffic logs, and geo-tagged tweets. 
% A huge amount of data with location information is called {\it big spatial data}.
We often analyze spatial data in many applications such as traffic control~\cite{wei2018intellilight}, human mobility analysis~\cite{pang2020intercity}, disaster surveillance~\cite{julian2019distributed}, mobility service~\cite{pan2019deep}, and route and point-of-interest recommendation~\cite{liu2019geo,sasaki2020sequenced}.
For efficient analysis of spatial data, it is essential to use distributed parallel processing systems for spatial data such as Sedona~\cite{sedona}, SpatialHadoop~\cite{spatialhadoop}, and others~\cite{sevim2021brief, aji2013hadoop,xie2016simba,tang2016locationspark,sasaki2021survey}.

% The volume of spatial data increases drastically due to the proliferation of the Internet of Things, so the importance of
% distributed parallel processing systems for spatial data such as Sedona~\cite{sedona} and SpatialHadoop~\cite{spatialhadoop} are increasing. 
% Deep reinforcement learning (DRL) has been attracting a lot of attention among both academic and industrial researchers, and achieved great success in many applications such as game, robotics, recommendation, and system optimization.
% One of the recent important topics on DRL is {\it big spatial data}.
% In big spatial data, DRL also has been studied on several applications such as traffic control~\cite{wei2018intellilight}, human mobility analysis~\cite{pang2020intercity}, disaster surveillance~\cite{julian2019distributed}, mobility service~\cite{pan2019deep}, and PoI recommendation~\cite{liu2019geo}.
% These works effectively discover new knowledge and accurately infer targets on their tasks.
% Since the volume of spatial data increases drastically due to proliferation of Internet of Things, the importance of DRL for big spatial data is increasing in several decades.
%Since a Gartner report \cite{gartner} mentions that the number of IoT devices is expected to grow to 20 billion by 2025, the trend will continue for decades.

\smallskip
\noindent
{\bf Motivation.}
% Although big spatial data is useful in many applications, it is inherently noisy and inaccurate because communication conditions and sensor qualities can be bad.
% In addition, because multiple organizations collect spatial data independently, spatial data are not integrated.
% Therefore, we should preprocess the big spatial data to remove/fix their noisy and inaccurate features and join data. 
%, and thus we need systems for efficiently managing big spatial data.
%Distributed parallel processing systems such as Spark~\cite{spark-hp} and Hadoop~\cite{hadoop-hp} are helpful to preprocess big data efficiently, but such common systems do not support efficient spatial data processing.
One of the fundamental functions of such systems is {\it spatial data partitioning}, which divides the whole area into sub-area (called partitions) and assigns data in the same partitions to computers for efficient parallel-processing~\cite{eldawy2015spatial}.
Existing spatial data partitioning methods focus only on data distributions to balance the computation costs among computers. 
However, it is difficult to obtain partitions that are optimized for data distributions, user queries, and computation frameworks.

{\it Deep reinforcement learning} (DRL for short) approach is effective for system optimization, including data partitioning~\cite{10.1145/3318464.3389704}. 
To our knowledge, there are no studies on DRL-based spatial data partitioning.
Since spatial data partitioning directly affects the performance of data processing, it is worth studying DRL-based spatial data partitioning to accelerate the processing of big spatial data.

\smallskip
\noindent
{\bf Contribution.}
We study spatial data partitioning with deep reinforcement learning. We have three main contributions; problem formulation, algorithm, and experiment.
% We have three main contributions: (1) formalizing spatial data partitioning problem in the context of reinforcement learning, (2) developing a novel learning algorithm for spatial data partitioning, and (3) evaluating our method using Apache Sedona with real-world data.

First, we formalize a spatial data partitioning problem in the context of reinforcement learning.
%This formulation defines state, action, and reward for spatial data partitioning.
The characteristics of our problem are to consider not only distributions of spatial data but also the computation environment (e.g., a computing system) and workloads (e.g., the set of queries).
This problem formulation aims at pursuing effective spatial data partitioning for each user.

Second, to address our spatial data partitioning problem, we develop a novel learning algorithm that efficiently explores actions for finding optimal partitions.
Our algorithm has a $2$-phase learning strategy consisting of pre-training and main-training~\cite{DQfD}.
In the pre-training phase, our algorithm trains a model by using pre-collected transitions (i.e., training data) based on existing spatial partitioning algorithms to avoid selecting less-effective actions during the main-training.
In the main-training phase, it searches for optimal partitions and trains the model using transitions (i.e., sequences of actions). %computed from actions.
Our framework is extended to (1) improve the effectiveness by using preparing demos of partitioning-related transitions and (2) accelerate learning by reducing action space and pruning run time measurements based on the characteristics of spatial data processing.

%Our learning framework combines deep Q-network~\cite{DQfD} with imitation learning~\cite{imitation_learning} to effectively train our model. 

Finally, we evaluate our method using Apache Sedona, the state-of-the-art distributed parallel processing system for big spatial data, with real-world datasets.
%Our method finds effective partitions that accelerate given workloads.
%, and the partitioning is completely different from the data-centric methods.
We validate that the DRL-based approach is highly effective in accelerating big spatial data processing.

% We summarize our contributions as follows:
% \begin{itemize}
%     \item Problem formulation: We first study spatial data partitioning from the context of reinforcement learning.
%     \item Novel algorithm: We develop an efficient and effective algorithm for deep reinforcement learning.
%     \item Experimental study: Through experimental study using Apatch Sedona and real-world datasets, we evaluate the effectiveness of our method. Our method reduces a workload run time consisting of distance join up to 59.4 \%.
% \end{itemize}

\smallskip
\noindent
{\bf Reproducibility.}
We open our source codes, datasets, and workloads that we used in an experimental study at Github https://github.
com/OnizukaLab/Spatial-Data-Partitioning-using-DRL.

% \smallskip
% \noindent
% {\bf Organization.}
% The rest of this paper is organized as follows. Section~\ref{prelininary} describes preliminaries, and Section~\ref{relatedwork} reviews existing works related to spatial partitioning and deep reinforcement learning. Section~\ref{problem-formulation} explains our problem formulation.
% %knowledge and Section~\ref{sec:4_graph_based_search} provides the problem settings. 
% %\ref{sec:4_graph_based_search} defines the problem of similarity search on computational notebooks and how to compute similarity.
% Section~\ref{learninglaogirhtm} presents our learning algorithm. 
% %To deal with the problems, we show the proposed fast searching algorithm in Section~\ref{sec:6_proposal}. 
% %Section~\ref{sec:6_proposal} describes how our method searches with subgraph matching.
% %Section~\ref{sec:6_proposal} presents optimization techniques of our search method.
% Section~\ref{experimentalstudy} shows the experimental study, Section~\ref{conclusion} concludes this paper with open challenges.

% We summarize our contributions as follows:
% \begin{itemize}
%     \item We first formulate the spatial data partitioning by the contexts of reinforcement learning.
%     \item We develop a learning algorithm that combines deep Q network and imitation learning.
%     \item We evaluate our approach to show its effectiveness.
% \end{itemize}

\section{Preliminaries} %0.5page
\label{prelininary}

We explain spatial data partitioning and deep reinforcement learning as preliminaries.

\subsection{Spatial data partitioning}
Spatial data partitioning is a function that determines how the area divides into partitions.
We define spatial data partitioning as follows:
\begin{definition}
Given a spatial dataset $\mathcal{D}$, spatial data partitioning is a task to divide the whole area in $\mathcal{D}$ into partitions and then assign sets of data in the same partitions to computers.
We denote the set of partitions by $\mathcal{P}$.
$\mathcal{P}$ needs to satisfy that any partitions $P \in \mathcal{P}$ are distinct and it covers the whole area in $\mathcal{D}$.
\end{definition}

Distributed parallel processing systems compose of multiple computers. 
The sets of data in partitions are assigned to computers and each computer parallelly processes its assigned data.
If computers need to access data that are held by other computers, they transfer their data (which is called shuffle), and finally, the processing results are aggregated to answer users.

Spatial data partitioning has two main objectives.
First, since a set of data that are spatially close to each other is often accessed together, the partitions should reduce the communication costs between computers.
Second, since computers parallelly process their data, partitions should have an equal amount of data to balance the burden of data processing.
These requirements are essential to scale out to multiple computers. 
%Regardless of the underlying architecture, e.g., disk-based or memory-based, data partitioning is essential to scale out to multiple computers. 

% \smallskip
% \noindent
% {\bf Related work on spatial data partitioning.}
% We review some typical algorithms for spatial data partitioning that are often implemented in distributed parallel processing systems~\cite{eldawy2015spatial,sasaki2021survey}. 
% Uniform grid is the simplest method that divides the area into equal-sized partitions.
% Quad-tree and KDB-tree are methods that repeatedly divide the partition that includes the largest number of data (initially starts a single partition covering whole area). 
% Quad-tree divides the partition into equal-sized four partitions, while KDB-tree divides the partition into two partitions so that the two partitions have the same number of data.
% Existing algorithms only focus on spatial data distributions without considering operators and computation environments.

\subsection{Deep reinforcement learning}
Reinforcement learning \cite{reinforcement_learning} is a process in which an agent repeatedly interacts with a Markov decision process to autonomously acquire information that serves as a supervisory signal while deriving optimal strategies. 
The agent selects {\it action} $a_i$ on {\it state} $s_i$ by {\it policy} $\pi(s_i,a_i)$, and then observes {\it reward} $r(s_i,a_i)$ and obtains the next state $s_{i+1}$ caused by $a_i$.
It repeats this process by selecting $a_1$ on $s_0$ until the state $s_n$ satisfies given conditions.
We define $\mathcal{A}$ and $\mathcal{S}$ as the action and state spaces, respectively.
A block of processes from $a_1$ to $a_n$ is called {\it episode}, and a four-pair $(s_t, a_t, r_t, s_{t+1})$ in an episode at step $t$ is called {\it transition}. 
% Such a trial and a sequence of actions $\langle a_1, \ldots , a_n \rangle$ are called {\it episode} and {\it transition}, respectively.
% The state also changes from $s_0$ to $s_{n}$ sequentially.

%In state $s_t$ at time $t$, the agent selects action $a_t$ according to the policy $\pi$, transitions to the next state $s_{t+1}$, and observes reward $r_{t}$. 
The goal of reinforcement learning is to maximize the cumulative reward $\Sigma_{k=0}^n r(s_{t+k},a_{t+k})$ at state $s_t$. 
It considers the action-value function $Q_{\pi}(s,a)=E[\Sigma_{k=0}^{\infty}\gamma^kr_{t+k}|s_t=s,a_t=a]$, which is the expected cumulative reward when the agent follows policy $\pi$.
%Therefore, we need to find the optimal policy $\pi^{*}$ that maximizes the expected cumulative reward. 
If we find the optimal function $Q^{*}$ to maximize the expected cumulative reward, we can derive the optimal policy as $\pi^{*}(s)=$ arg~max$_{a\in \mathcal{A}}Q^{*}(s,a)$ for $\forall s \in \mathcal{S}$.

{\it Deep reinforcement learning} aims at approximating the action-value function by using deep neural networks.
It is often difficult to maintain the function for any pairs of actions and states if their patterns are tremendous.
When we train the deep neural network models with tremendous patterns, the agent can select an optimal action on each state following these models.

\smallskip
\noindent
{\bf Q-learning and Deep Q-network.} 
Q-learning~\cite{q-learning}, a typical reinforcement learning method, selects a high-value action in terms of the action-value function $Q$. 
By repeatedly observing the state and reward, the optimal action-value function is updated using the temporal difference error between the  outputs of $Q$, which are called $Q$-values, in the next and current states.

\begin{equation}
    \label{q-learning}
    \begin{split}
        Q(s_t,a_t){\leftarrow}&Q(s_t,a_t)+\\
        &\!\!\!\!\!\!\alpha\{r_{t+1}+\gamma\max_{a\in \mathcal{A}}Q(s_{t+1},a_{t+1})-Q(s_t,a_t)\}
    \end{split}
\end{equation}
If the action and/or state spaces are large or continuous, the number of combinations becomes very large, making it difficult to maintain $Q$-values in the table.

Deep Q-Network (DQN), which is an extension of Q-learning, approximates the function $Q$ and the policy function in Q-network.
%to construct a network that outputs the Q-values of actions with the states as inputs.
Similar to Q-learning, DQN aims to make $Q(s_t,a_t)$ close to $r_t+\gamma\max Q(s_{t+1},a_{t+1})$, so the Q-network is trained using the error function $L$ and the current $Q$-value as follows:
%, with the latter as the target. 

\begin{equation}
    \label{dqn_loss}
    L = \sum_{(s_t, a_t, r_t, s_{t+1}) \in B}(r_t+\gamma \max Q(s_{t+1},a_{t+1};\theta)-Q(s_t,a_t;\theta))^2,
\end{equation}
\noindent
where $\theta$ and $B$ are a deep neural network model and a set of past transitions, respectively.

\section{Related work}
\label{relatedwork}
We review existing works related to spatial data partitioning and system optimization with deep learning.
To the best of our knowledge, there are no deep learning-based methods for computing spatial data partitioning. 

%\smallskip
\noindent
{\bf Spatial data partitioning}
We review some typical algorithms for spatial data partitioning. 
Uniform grid is the simplest method that divides the area into equal-sized partitions.
Quad-tree~\cite{quad-tree} and KDB-tree~\cite{kdb-tree} are methods that repeatedly divide the partition that includes the largest number of data (initially starts a single partition covering the whole area). 
Quad-tree divides the partition into equal-sized four partitions, while KDB-tree divides the partition into two partitions so that the two partitions have the same amount of data.
Vu~\cite{vu2021incremental} proposes a method for incremental updates of spatial partitions which utilize estimation of query execution time.
Vu et al.~\cite{10.1145/3402126} proposed a method that selects optimal spatial partitioning methods (e.g., KDB-tree and Quad-tree) by using DRL. This method just selects methods among the given ones, instead that it does not aim at computing partitions.
Aly et al.~\cite{aly2015aqwa} proposed a spatial partitioning method that uses a given set of range and kNN queries. This aims to reduce the communication between computers, but it can use only range and kNN queries instead of spatial join, so it is not applicable to general settings that include many query types.

Hilprecht et al.~\cite{10.1145/3318464.3389704} proposed a DRL-based hash partitioning, but it does not focus on spatial partitioning. It is well known that hash partitioning is not effective for spatial queries. 

%Existing algorithms for spatial partitioning do not use only focus on spatial data distributions without considering operators and computation environments.

% We here note that these partitioning algorithms have similar procedures to spatial data {\it indexing} methods, so they have the same name of indexing methods.
% However, their purposes are different each other: spatial data partitioning assumes distributed environments to effectively reduce the communication costs between machines and balance the data processing burdens, while spatial data indexing aims efficient data access in a single machine. 

% Distributed parallel processing systems often support spatial data partitioning~\cite{eldawy2015spatial,sasaki2021survey}, for example Hadoop-GIS~\cite{aji2013hadoop}, Spatial Hadoop \cite{eldawy2015spatialhadoop}, Simba~\cite{xie2016simba}, Magellan~\cite{magellan}, LocationSpark~\cite{tang2016locationspark}, GeoSpark (known as Apache Sedona)~\cite{yu2015geospark}, and SpatialSpark~\cite{you2015large}.
% Therefore, effective spatial data partitioning has the large impact to big spatial data processing.

\smallskip
\noindent
{\bf System optimization with deep learning}.
% DRL algorithms including DQN have evolved over the years.
% Many techniques have been proposed on DRLs to enhance the accuracy and reduce train time, for example double DQN~\cite{DoubleQlearning}, dueling network~\cite{DuelingNetwork}, noisy network~\cite{NoisyNetwork}, prioritized experience replay~\cite{PrioritizedExperienceReplay}, categorical DQN~\cite{CategoricalDQN}, multi-step learning~\cite{deasis2018multistep}.
% Rainbow~\cite{2017rainbow} composes all techniques and improve the performance.
% One of the main challenges to apply DRLs is what techniques we adopt in our algorithm. 
% In our study, we use prioritized experience replay and multi-step learning, which largely contribute to the performance. The further optimization remains the future work.
% multi step learning~\cite{deasis2018multistep}, double Q learning~\cite{DoubleQlearning}, dueling network~\cite{DuelingNetwork}, and rainbow~\cite{2017rainbow}.
% One of the main challenges to apply DRLs is that how combines the existing methods.
% In our study, we use 
Spatial data partitioning is one of the tasks in system optimization, so we review system optimization techniques using deep learning~\cite{li2017deep}. 
% Reinforcement learning is expanding its application to more complex problems, such as conquering imperfect information games and intelligent systems in real space, such as poker and automated driving~\cite{poker,autonomous_vehicle2019}.
% For example, it has been reported that AI has defeated professional players in poker \cite{poker}, applied to areas such as automated driving technology \cite{autonomous_vehicle2019}\cite{toromanoff2020endtoend}, and computer cluster resource management \cite{resource_manegment}. 
%Spatial data partitioning is one of the tasks in system optimization.
%DRL-based approaches have been studied to system optimization.
Spatial partitioning is similar to indexing techniques on spatial/multi-dimensional data, and many learning indexing techniques are developed, such as Flood~\cite{nathan2020learning}, Tsunami~\cite{ding2020tsunami}, LISA~\cite{li2020lisa}, Qd-tree~\cite{yang2020qd}, RLR-tree~\cite{https://doi.org/10.48550/arxiv.2103.04541}, and RSMI~\cite{qi2020effectively}.
They build data blocks (i.e., the set of data) and construct index structures to accelerate spatial query processing on a single machine.
Since they do not assume spatial partitioning, their blocks are either unsophisticated or inapplicable for spatial partitioning, for example, Flood divides data into blocks based on the data distribution on a single axis and the blocks of Qd-tree are not distinct.
% Among them, Flood, Tsunami, and LISA, their data blocks can be considered as spatial partitions because data are not overlapped across blocks. However, since they do not assume distributed processing, their ways for partitions are not sophisticated, for example, Flood divides data into blocks based on the data distribution on a single axis and LISA divides data into blocks of grid cells that have the equal number of data.
% The blocks of Qd-tree cannot be used for spatial partitions because the generated blocks are overlapped.
Lan et al.~\cite{10.1145/3340531.3412106} proposed a DRL-based index recommendation. 
Vu et al.\cite{vu2021learned} develop a query optimizer for spatial join based on neural networks, which aims to estimate the cardinality and select join algorithms. These works do not aim to divide the dataset into data blocks  
%Consequently, there are no DRL methods for spatial partitioning.
%These methods cannot apply to spatial partitioning because they do not aim to partition a given spatial data.
%Gu et al.~\cite{https://doi.org/10.48550/arxiv.2103.04541} proposed a DRL-based R-tree indexing.

\section{Problem Formulation}
\label{problem-formulation}
We formulate spatial data partitioning as a new reinforcement learning problem. 
%We summarize the symbols that we use in this paper.
First, we define our problem as follows:

% \begin{table}[t]
%     \centering
%     \caption{Summary of notations}
%     \label{hyperparameter}
%     \begin{tabular}{ll} \hline
%         \bf{Symbol} & \bf{Meaning} \\\hline
%         $\mathcal{D}$ & Dataset \\
%         $W$ & a set of worklords \\
%         $\mathcal{P}$ & set of partitions \\
%         $\mathcal{P}_b$ & best set of partitions during training \\
%         $\mathcal{P}_e$ & set of partitions at the current episode \\
%         $s_{(i,j)}$ & state of grid cell $(i,j)$\\
%         $a$ & action \\
%         $\mathcal{A}$ & set of all actions \\
%         $r$ & reword \\
%         \hline
%     \end{tabular}
% \end{table}

\smallskip
\noindent\textbf{Problem:}
Given a spatial dataset $\mathcal{D}$, a workload $W$ (i.e., the set of queries and their frequencies), and a computation environment (i.e., system and the number of computers), the goal is to find the set of partitions $\mathcal{P}$ that minimizes the run time of the workload for $\mathcal{D}$ on the computation environment.

\smallskip
In the following, we define the initial setup, state, action, and reward for this problem, respectively.

% The definition of state assumes that data distribution of input dataset is clear.
% It is also assumed that the distance threshold and the percentage of Distance Join in the workload are known.
% We target the $\epsilon$ Distance Join Query ($\epsilon$DJQ) for acceleration, which is defined as follows.

% \noindent\textbf{Distance Join Query}
% Let $\mathbb{P}=\{p_0,p_1,\ldots,p_{n-1}\}$ and $\mathbb{Q}=\{q_0,q_1,\ldots,q_{m-1}\}$ be two set of points in $E^d$, and a distance threshold $\epsilon \in \mathbb{R}^+$. The results of $\epsilon$Distance Join Query ($\epsilon$DJQ) of P and Q, 
% $(\epsilon DJQ(P, Q, \epsilon) \subseteq P \times Q)$, is a set which contains all the possible diﬀerent pairs of points from P×Q that have a distance of each other smaller than, or equal to ε:
% $\epsilon DJQ(P,Q,\epsilon)=\{(p_i,q_j)\in P \times Q:dist(p_i,q_j) \leq \epsilon\}$

\smallskip
\noindent\textbf{Initial setup.}
We form a map that covers the locations of the given data in a two-dimensional space as an environment for a DRL agent.
%, as shown on the left in Figure \ref{environment}.
We divide this map into a set of uniform grid cells. 
The agent constructs partitions by adding new boundaries on each grid line.

\smallskip
\noindent\textbf{State:}
The state is obtained from the current partitions and data distribution.
We define that state $s$ is a set of status $s_{(i,j)}$ for each grid cell $(i,j)$ as follows:
\begin{equation}
  s_{(i,j)}=(h_\mathit{(i,j)}, v_\mathit{(i,j)}, p_\mathit{(i,j)}),
\end{equation}
where $i$ and $j$ specify the upper-left coordinate of the grid cell. 
$h\in\{0,1\}$ and $v\in\{0,1\}$ denote the absence/presence of horizontal and vertical boundaries on the top and left sides of the cell, respectively. 
$p_{\mathit{(i,j)}}$ denotes the ratio of the numbers of data in the cell and partition. We define $p_{\mathit{(i,j)}}=\frac{|\mathcal{D}_{\mathit{(i,j)}}| \cdot |P_{\mathit{(i,j)}}|}{|\mathcal{D}|^2}$, where  $|\mathcal{D}|$, $|\mathcal{D}_{\mathit{(i,j)}}|$, and $|P_{\mathit{(i,j)}}|$ are the numbers of data in the given dataset $\mathcal{D}$, in a cell $(i,j)$, and in a partition that covers $(i,j)$, respectively. 
%is the product of the percentages of the numbers of data in cell $(i,j)$ over all data and in the partition that $(i,j)$ belongs over all the data.

\smallskip
\noindent\textbf{Action.}
The action $a\in \mathcal{A}$ is an operation to add a new boundary line. We define the action as follows:
\begin{equation}
a=(i, j, \mathit{dir}),
\end{equation}
where $i$ and $j$ are the coordinates of the starting point of the action and $\mathit{dir} \in\{\mathit{right}, \mathit{down}\}$ specifies the direction of the boundary from $(i,j)$, i.e., rightward or downward. 
To avoid invalid actions, the starting points need to be satisfied either of conditions; (1) $\mathit{dir} = \mathit{right}$, $h(i,j) =0$, and $v(i,j) =1$ or (2) $\mathit{dir} = \mathit{down}$, $h(i,j) =1$, and $v(i,j) =0$.

\smallskip
\noindent\textbf{Reward.}
The reward $r$ is computed based on the run time $C(\mathcal{P},w)$ for an operator $w$ on a set of partition $\mathcal{P}$ to the spatial dataset in the computation environment.
It enables to maximizing the reward $r$ to optimize execution in the given computing environment.
The reward $r$ is evaluated as follows:

\begin{equation}
    r = \left(\sum_{w_i \in W}{f_i}\cdot\frac{C(\mathcal{P}_\mathit{b},w_i)}{C(\mathcal{P}_\mathit{e},w_i)}\right)^2,
\end{equation}
where $w_i$, $f_i$, $\mathcal{P}_{\mathit{e}}$, and $\mathcal{P}_b$ indicate a query in the workload, a frequency of operator $w_i$, a set of partitions at the end of the episode, and the best set of partitions we found during training, respectively.
This equation evaluates a relative to the best run time up to that on the partitions at the current episode.  
The value is squared to emphasize the difference between the run time on $\mathcal{P}_b$ and $\mathcal{P}_e$.
If $r>1$, $\mathcal{P}_e$ is better than $\mathcal{P}_b$.
%, and thus $\mathcal{P}_b$ is replaced by $\mathcal{P}_e$ from the next episode.
Note that computing all queries in the workload is costly compared with other procedures such as training deep neural networks.
%increase the reward value for $r>1$ and decrease the reward value for $r<1$.
%The median value from multiple measurements is used to calculate the reward.

% The initial value of $P_{\mathit{best}}$ is a KDB-tree, the best among existing algorithms for most data distributions. $P_{\mathit{best}}$ is updated whenever a partition with $r > 1$ is obtained during training.

% \smallskip
% In general, the processing will be shorter if the number of data in each partition is equal and the data traffic during the computation period is small for pre-processing in the workload.

\begin{figure*}[t]
    \centering
    \includegraphics[width=0.7\linewidth]{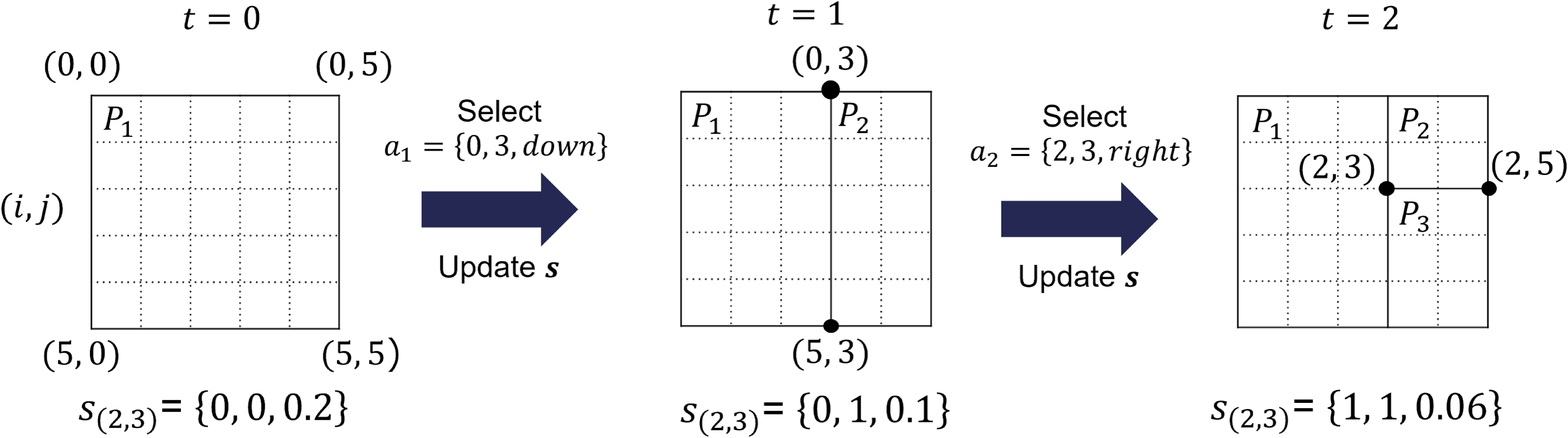}
    \caption{Example of actions and states: the agent selects actions $a_t$ to add new boundaries depending on the state $s_t$.}
    \label{environment}
\end{figure*}

\smallskip
\noindent\textbf{Example.}
Figure \ref{environment} illustrates an example of states and actions for obtaining three partitions.
Initially, there are no boundaries (i.e., a single partition).
In this example, the agent takes $a_1 =(0,3,down)$ and $a_2 =(2,3,right)$.
The status of cell $s_{(2,3)}$ changes by adding boundary lines. $p_{(2,3)}$ becomes smaller because the percentage of the number of data in the partition becomes smaller instead of that in the cell is constant.
We run the workload and obtain the reward for the episode after computing three partitions. 
%The cell corresponding to $s_{(2,3)}$ changes $h$ and $v$ from $0$ to $1$ by adding a boundary line that passes through its upper or left side. The $p_\mathit{(i,j)}$ is the product of the percentage of data in the cell (blue) and in the partition to which cell belongs (orange).

\section{Learning Algorithm}
\label{learninglaogirhtm}
This section proposes a learning algorithm for our problem formulated in Section~\ref{problem-formulation}.  
We have two challenges to solve our problem.
First, the action and state spaces are very large to find optimal partitions.
Common DRL algorithms start the random selection of actions.
We should efficiently explore actions by capturing the characteristics of spatial data partitioning.
%So, we need an effective way that 
%Precise partitions require a more refined grid size, while the state and action spaces are larger.
Second, our problem takes a large training time because of measuring the run time of workload at the end of each episode to compute rewards.
%Because the run time is typically very large compared with updating deep neural networks, it takes a large time for training the deep neural networks. 
For efficient training, we should reduce the run time while keeping the effectiveness of training.
%, which increases in proportion to the data size and query distance,

To address the above challenges, we design a learning strategy consisting of (1) pre-training using {\it demo data} which is a set of pre-collected transitions and (2) main-training.
Our learning framework trains our model by effectively combining DQN with imitation learning based on our strategy.
Sections~\ref{strategy} and~\ref{framework} present our learning strategy and learning framework, respectively.
Section \ref{algorithm} presents our algorithm.

\begin{comment}
We design a novel learning algorithm to efficiently train a model that can utilize effective actions during training. 
%combines DQN with imitation learning.
%Imitation learning aims to mimic pre-collected transitions, called {\it demo data}, to avoid selecting less-effective actions.
%to effectively uses the superior experience acquired by agents.
%for efficient partitioning learning.
Our algorithm has a $2$-phase learning strategy consisting of (1) pre-training using {\it demo data} and (2) main-training.
Demo data is a set of pre-collected transitions, and the pre-training phase aims that our model selects actions in the demo data, for avoiding selecting less-effective actions during the main-training.
In the main-training phase, our algorithm searches for optimal partitions and trains our model by using demo data and transitions computed from the actions of agent.
Our learning framework combines DQN with imitation learning to effectively explore actions by using demo data.
%In these training phases, it trains a new learning model that is effective to find the optimal partitions. 
%The demo data is a pre-defined transition and the agent data are the transitions that the agent has explored during the learning.
We present our learning strategy and framework in Sections~\ref{strategy} and~\ref{model}, respectively.
\end{comment}

\subsection{Learning strategy}
\label{strategy}

Our algorithm is based on the following ideas:
%In order to promote efficient and effective learning, 
\begin{itemize}
    \item   Pre-training by effective demo data:
            Our algorithm builds a model that follows existing algorithms for spatial data partitioning.
            This model avoids selecting less-effective actions at early episodes of the main-training phase.
    \item   Effective new action choice:
            We add a new action that makes a boundary close to the best partitions.
            It supports exploring effective actions. 
            %They work well for finding better partitions than the current best ones.
    \item   Pruning run time measurement:
            Our algorithm prunes ineffective executions of the workload to reduce the training time. 
\end{itemize}
These ideas are specialized for our problem formulation, and thus all of them are not applied to the DRL approach yet. We explain pre-training and main-training phases.

\smallskip
\noindent
\textbf{Pre-training.}
In the pre-training phase, we train our model using only demo data so that the agent similarly selects actions in demo data.
We first collect transitions from actions based on existing partitioning algorithms and then use the transitions as demo data.
It makes the initial values of the agent's actions appropriate for the distribution of a given dataset before the main-training.
Therefore, this pre-training phase drastically reduces random actions in the early episodes of the learning processes, which is essentially an exhaustive trial-and-error process.

The selection of demo data may not obtain exactly the same partition computed from existing methods because the boundaries of our partitions are limited on pre-defined grid cells. 
We select the closest boundaries to those of existing methods. 
%The Q-network is trained using a mini-batch sampled from only this demo data.
Our strategy guarantees that the run time on the best partitions obtained by our algorithm is at least the (almost) best performance among existing methods by choosing the most effective method for generating demo data.

\smallskip
\noindent
\textbf{Main-training.}
In the main-training phase, we search for actions that construct a set of partitions better than those generated by the pre-collected demo data.
The basic procedures of the main-training phase are that the agent repeatedly selects its actions probabilistically by using an $\epsilon$-greedy method~\cite{thrun2000reinforcement}, and then when the number of partitions reaches the number of machines, it distributes data into computers according to the partitions, executes the workload, and computes a reward. 
Our algorithm repeatedly runs these procedures until reaching the given number of episodes. 
%We call the transitions computed by agents {\it agent data}.

Our algorithm adds new action choices to the $\epsilon$-greedy method.
In the common $\epsilon$-greedy method, the agent selects the best action approximated by a model or a random action. 
We add actions of one {\it grid-shift} from the best action to the $\epsilon$-greedy method.
In spatial data partitioning, better partitions than the current best ones are often to be found near the current best partitions, and thus such one grid-shift actions help to search for the optimal partitions effectively. 
We can find effective partitions efficiently by intensively exploring actions near the approximated best action.
%By focusing the exploration around the actions that construct the best partitions, we can explore for effective partitions efficiently.

We reduce the training time by terminating the workload execution for partitions that are clearly less-effective. 
%The workload run time is compared with that for $P_{b}$ and 
Specifically, the workload execution stops whenever the execution time of one of the queries in the workload exceeds a certain level compared with that on $\mathcal{P}_{b}$.
In this case, its reward $r$ is set to a predefined small value.
%This idea drastically reduces the training time.
%limit less than or equal to the value calculated backward from the set time.

%This main-training phase can efficiently explore for better partitions with reducing training time.

\subsection{Learning framework}
\label{framework}
Our learning framework consists of an agent, a learner of the network that approximates the action values of the agent, and two replay memories. These replay memories are called demo and agent memories, storing the demo data and other transitions, respectively.
The demo memory is used for imitation learning.
We note that it uses only demo memory to train deep neural networks in the pre-training phase.

% \begin{figure}[t]
%     \centering
%     \includegraphics[width=\linewidth]{figure/learning_model.eps}
%     \caption{Learning framework}
%     \label{learning_model}
% \end{figure}

In the main-training phase, our framework adds transitions that construct better partitions than $\mathcal{P}_b$ to demo data. 
This contributes to exploring better actions by imitating the current best actions instead of the pre-collected actions.
More concretely, if $r > 1$, it adds the transitions to the demo data.
%because the partition of the current episode $P_e$ is better than $P_b$.
In addition, we add transitions with $r \leq 1$ to another memory because such data is effective to learn the wide variety of action values.  
% \hori{However, because of the additional loss to imitate the demo data, the action values other than demo data are different from the actual values. 
% Therefore, we add the transitions for $r < 1$ to another memory and train to modify the action value to its actual value, as in common DQN.}

We use two Q-networks, main network $\theta$ and target network $\theta'$, following existing works~\cite{DQN,gorilla}. The target network is used for stable learning. 
The learner trains the main network and periodically overwrites $\theta'$ by $\theta$.
It also updates priorities for transitions in memories using prioritized experience replay~\cite{PrioritizedExperienceReplay}, which is effective to sample data for training the Q-network. 
%The priorities are computed based on temporal difference error of PER to achieve effective data sampling for training.
The input and output dimensions of the Q-network are the size of the state and action spaces, respectively. 
Mini-batch data sampled from both memories according to the priorities and the ratio $\rho$ that controls the importance of memories.
The imitation learning guarantees that the mini-batch data include effective transitions sampled from the demo memory, resulting in Q-networks that are trained to find better partitions.

For the loss function, we use three losses.
First, loss $L_N(\theta)$ is the temporal difference error considering the rewards of $n$ steps ahead using multi-step learning~\cite{deasis2018multistep} as follows:
%shown in Equation (\ref{multisteploss}). 

\begin{equation}
\begin{split}
    L_N(\theta) = \sum_{(s_i, a_i, r_i, s_{i+n})\in B}(r_i+\gamma^1r_{i+1}+...+\gamma^{n+1}r_{i+n-1}\\
    + arg~max_{a \in \mathcal{A}}Q_{\theta'}(s_{i+n},a_{i+n})-Q_\theta(s_i,a_i))^2
    \label{multisteploss}
\end{split}
\end{equation}
This loss accelerates the propagation of the reward, which is obtained only at the end of the episode, thereby improving learning efficiency.

Second, to imitate the actions in demo data, we use large margin classification loss $L_C(\theta)$~\cite{DQfD}, which is applied only to the demo data.
$L_C(\theta)$ is computed as follows:
\begin{equation}
    L_{C}(\theta)=\max_{a\in \mathcal{A}}[Q(s,a)+l(a_C,a)]-Q(s,a_C),
    \label{large_margin}
\end{equation}
where $a_C$ and $l(a_C,a)$ are an action in demo data and a function that outputs $0$ when $a=a_C$ and positive otherwise, respectively. 
$L_C(\theta)$ becomes zero if $a_C$ is selected as the best action, otherwise the loss becomes larger to imitate the demo data. 
This loss effectively accelerates imitations. 

Finally, we use the L2 regularization loss, $L_{l2}(\theta)$, for parameters $\gamma_i$ of main-network $\theta$ to suppress overfitting to a small number of demo data.
\begin{equation}
    \label{l2_norm}
    L_{l2}(\theta) = \sum_{\gamma_i \in \theta}{|\gamma_i|}^2
\end{equation}

We apply the three losses to loss $L(\theta)$ as follows:
\begin{equation}
    \label{our_loss}
    L(\theta) = \lambda_1 L_N(\theta) +  \lambda_2 L_C(\theta) + \lambda_3 L_{l2}(\theta),
\end{equation}
where $\lambda_1$, $\lambda_2$, and $\lambda_3$ are the weights for the corresponding losses.

% モデルの学習アルゴリズム
\begin{algorithm}[!t]
    \caption{Learning algorithm} \label{train:algorithm}
        \DontPrintSemicolon
            \SetKwInOut{Input}{input}
    	    \SetKwInOut{Output}{output}
    	    \SetKwInOut{Parameter}{parameter}
    	    \Input{Dataset $\mathcal{D}$, workload $W$, the number of computers $M$, demo data $B_\mathit{d}$, demo partitions $\mathcal{P}_d$}
    	    \Parameter{Update weight frequency $U$, demo ratio $\rho$, the number of episodes $e_{max}$}
    	    \Output{Best partitions $\mathcal{P}_b$}
    	    {\bf procedure} {\sc Pre-training}\\
    	    Initialize the main network $\theta$ and target network $\theta'$\;
            $B_\mathit{demo} \gets B_\mathit{d}$\;
            
            \For{$e=1, \ldots, e_{max}$}{
                Sample mini-batch $B$ from $B_\mathit{demo}$\;
                Train main network $\theta$ with $B$\;
                {\bf if} $(e \bmod U) = 0$ {\bf then} $\theta' \leftarrow \theta$\;
                Generate partitions $\mathcal{P}_e$ by agent actions\;
                {\bf if} $\mathcal{P}_e= \mathcal{P}_d$  {\bf then} {\bf break}
            }
            
            {\bf procedure} {\sc Main-training}\\
            $\mathit{step} \gets 0$, $\mathcal{P}_b \gets \mathcal{P}_d$, $B_\mathit{agent} \gets \varnothing$\;
            \For{$e=1, \ldots, e_{max}$}{
                Reset state $s_0$, $B_\mathit{tmp} \gets \varnothing$\;
                \For{$t=1, \ldots, M-1$}{
                    Choose action $a_t$ with our $\epsilon$-greedy method by $\theta$\;
                    Execute action $a_t$ and observe state $s_{t}$\;
                    \eIf{$t \neq M-1$}{
                        $B_\mathit{tmp} \gets B_\mathit{tmp} \cup \{(s_t, a_t, 0, s_{t+n})\}$\;
                    }
                    {
                        Distribute partitions $\mathcal{P}_e$ to computers\;
                        Run operators in $W$\;
                        $r_t = (\sum_{w_i \in W}{f_i}\cdot\frac{C(\mathcal{P}_b,w_i)}{C(\mathcal{P}_e,w_i)})^2$\;
                        $B_\mathit{tmp} \gets B_\mathit{tmp} \cup \{(s_t, a_t, r_t, s_{t+n})\}$\;
                        \eIf{$r > 1$}{
                            $B_\mathit{demo} \gets B_\mathit{demo} \cup B_\mathit{tmp}$\;
                            $\mathcal{P}_b \gets \mathcal{P}_e$\;
                        }{
                            $B_\mathit{agent} \gets B_\mathit{agent} \cup B_\mathit{tmp}$
                        }
                    }
                    Sample mini-batch $B$ from $B_\mathit{demo}$ and $B_\mathit{agent}$\;
                    Train main network $\theta$ with $B$\;
                    $\mathit{step} \gets \mathit{step}+1$\;
                    {\bf if} $(\mathit{step} \bmod U) = 0$ {\bf then} $\theta' \leftarrow \theta$
                }
            }
            %{\bf end procedure}
\end{algorithm}

\subsection{Algorithm}
\label{algorithm}

Algorithm \ref{train:algorithm} shows a pseudo-code of our learning algorithm.
In the pre-training phase, we store the pre-collected demo data to the demo memory $B_\mathit{demo}$ (line 3) and then train the main network $\theta$ using the mini-batch data sampled from demo data (lines 4--9). 
We repeat the training until the partition $\mathcal{P}_e$ constructed by the agent is the same as the partitions $\mathcal{P}_d$ that demo data constructs. 
%$J(\mathcal{P}_e,\mathcal{P}_d)$ is a function that identifies whether two partitions are the same or not.

In the main-training phase, the agent repeatedly collects transitions. We set the reward $r_t$ for each step $t$ in the episode to zero (line 18) and measure the run time of the workload when the number of steps reaches $M-1$ (i.e., the numbers of partitions and computers are the same) to compute the reward (lines 20--22). All transitions during the episode are stored in a temporary memory $B_\mathit{tmp}$ (line 23) and allocated to the demo memory $B_\mathit{demo}$ or agent memory $B_\mathit{agent}$ according to the reward $r_{M-1}$ (lines 24--28). 
We use mini-batch data sampled at a rate $\rho$ specified from $B_\mathit{demo}$ and $B_\mathit{agent}$ to train $\theta$ (lines 29--30).
$\theta$ overwrites the target network $\theta'$ follows update frequency $U$ (line 32). 
Our algorithm outputs the best partitions $\mathcal{P}_b$ found during training after finishing $e_{max}$ episodes.

\section{Experimental study} 
\label{experimentalstudy}
%We explain the results of an experimental evaluation of our method.
We designed the experiments to clarify the following questions:
(1) Is our method effective compared with existing methods?
(2) Does our method find partitions efficiently?
%(3) Is our method robust to data increase and workload changes?
%(Some additional explanations and experimental results appear in our supplementary file.)

\subsection{Experimental setting}
%We provide an overview of our experimental setup.
%
%
%\noindent
{\bf Computation environment.}
We used Apache Sedona, the state-of-the-art distributed and parallel processing system for big spatial data~\cite{sedona}.
We used nine computers with a Linux server with 32GB of memory and an Intel Celeron G4930T CPU@3.00GHz processor.
One computer was a master server and the others were workers (i.e., eight computers processed a given workload). 
Sedona provides spatial queries, such as kNN, range, and distance join, and partition methods, such as KDB-tree and R-tree.
We evaluated two cases with and without a local index, which supports efficient data access from the workers to their own local data.

To design our setting, we consider the limitations of Sedona; (a) range and kNN queries use hash partitioning instead of spatial partitioning, (b) R-tree partitioning cannot control the number of partitions, and (c) Polygon data is not handled on spatial join. The design of our experimental study is following to these limitations.

\begin{figure}[t]
    \centering
	\subfigure[OSM-US]{%
		\includegraphics[clip, width=0.4\hsize]{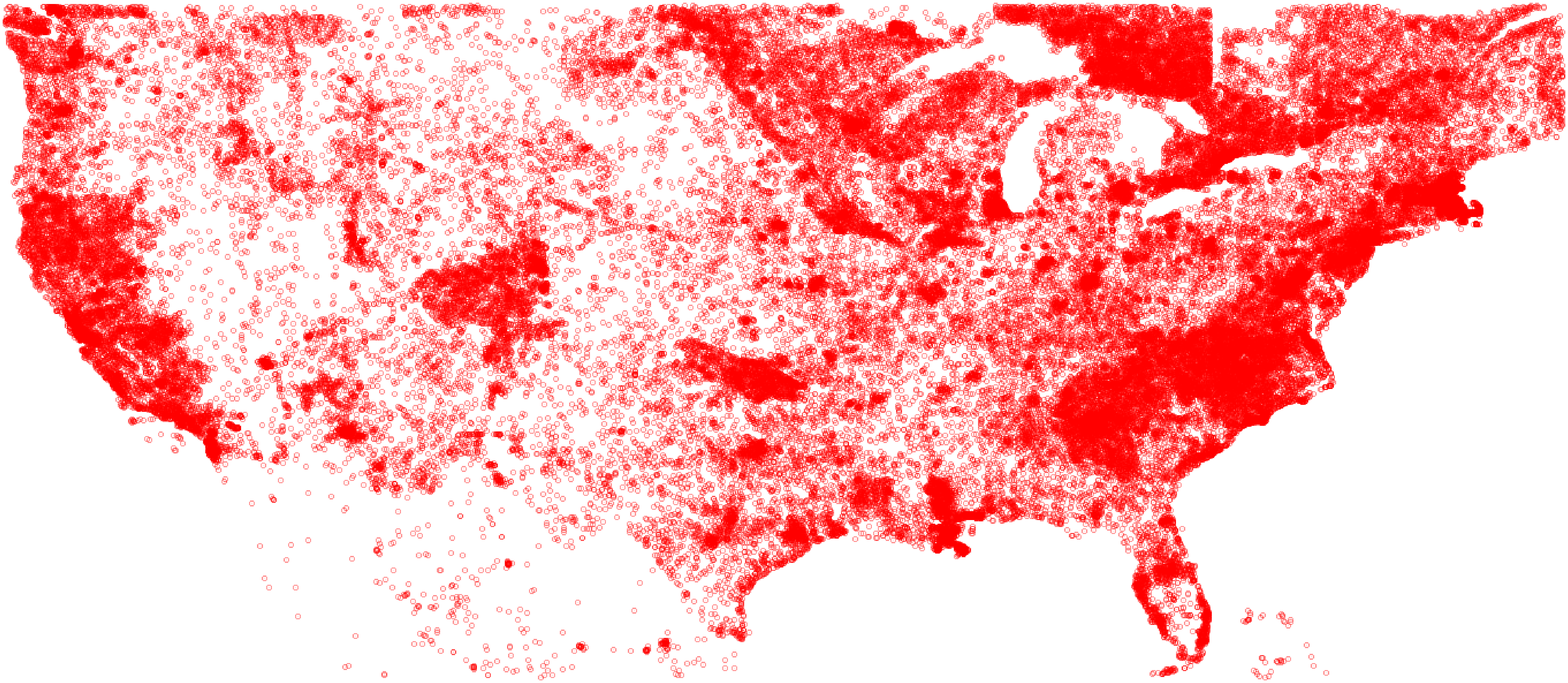}}%
	\subfigure[OSM-SA]{%
		\includegraphics[clip, width=0.3\hsize]{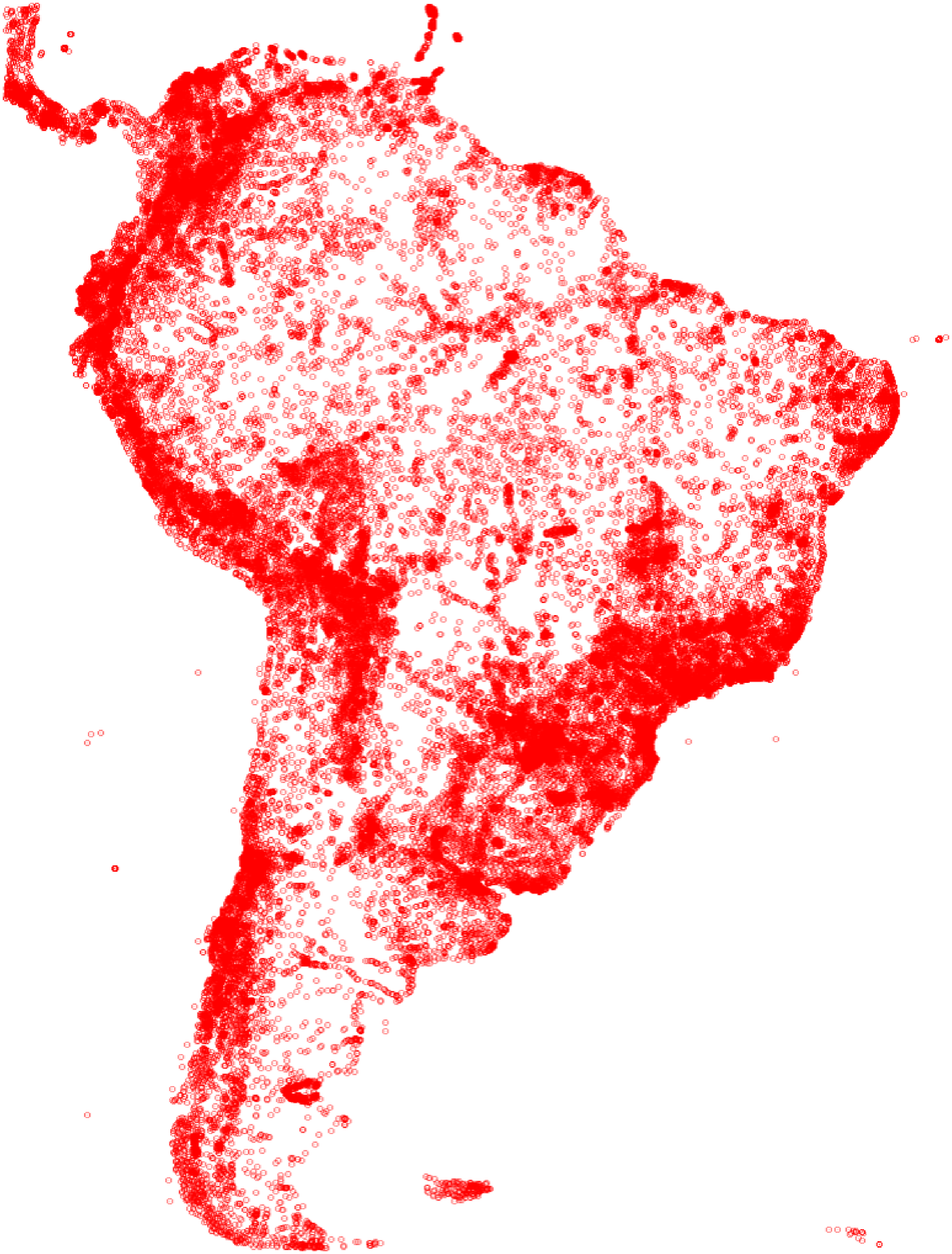}}%
    \subfigure[Imis]{%
		\includegraphics[clip, width=0.3\hsize]{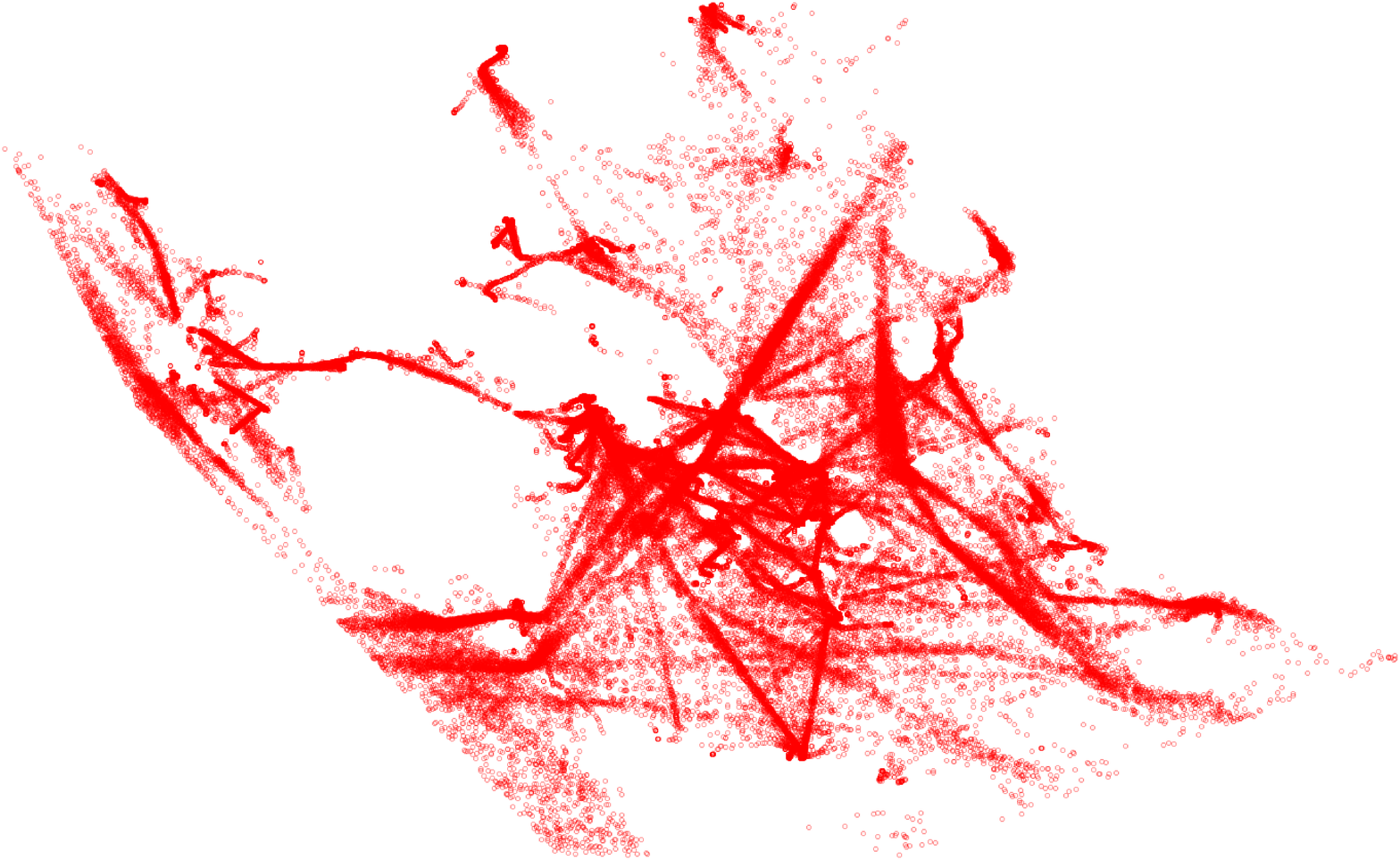}}%
	\\
	\caption{Data distributions}
	\label{data-distributions}
\end{figure}

\smallskip
\noindent
{\bf Datasets.}
We used point-of-interest data in two different areas, the United States (US) and South America (SA), extracted from OpenStreetMap\footnote{http://osm.db.in.tum.de/}, and Integrated Marine Information System (Imis)\footnote{https://chorochronos.datastories.org/?q=content/imis-3months}, which is.
For each dataset, we randomly sampled 100,000 point-of-interest from the original datasets.
%Figure~\ref{data-distributions} illustrates the distributions of datasets on OSM-US and OSM-SA.
Figure~\ref{data-distributions} shows data distributions. They have different area sizes and distributions. 

\begin{figure*}[t]
    \centering
	\subfigure[Uniform grid]{%
		\includegraphics[clip, width=0.3\hsize]{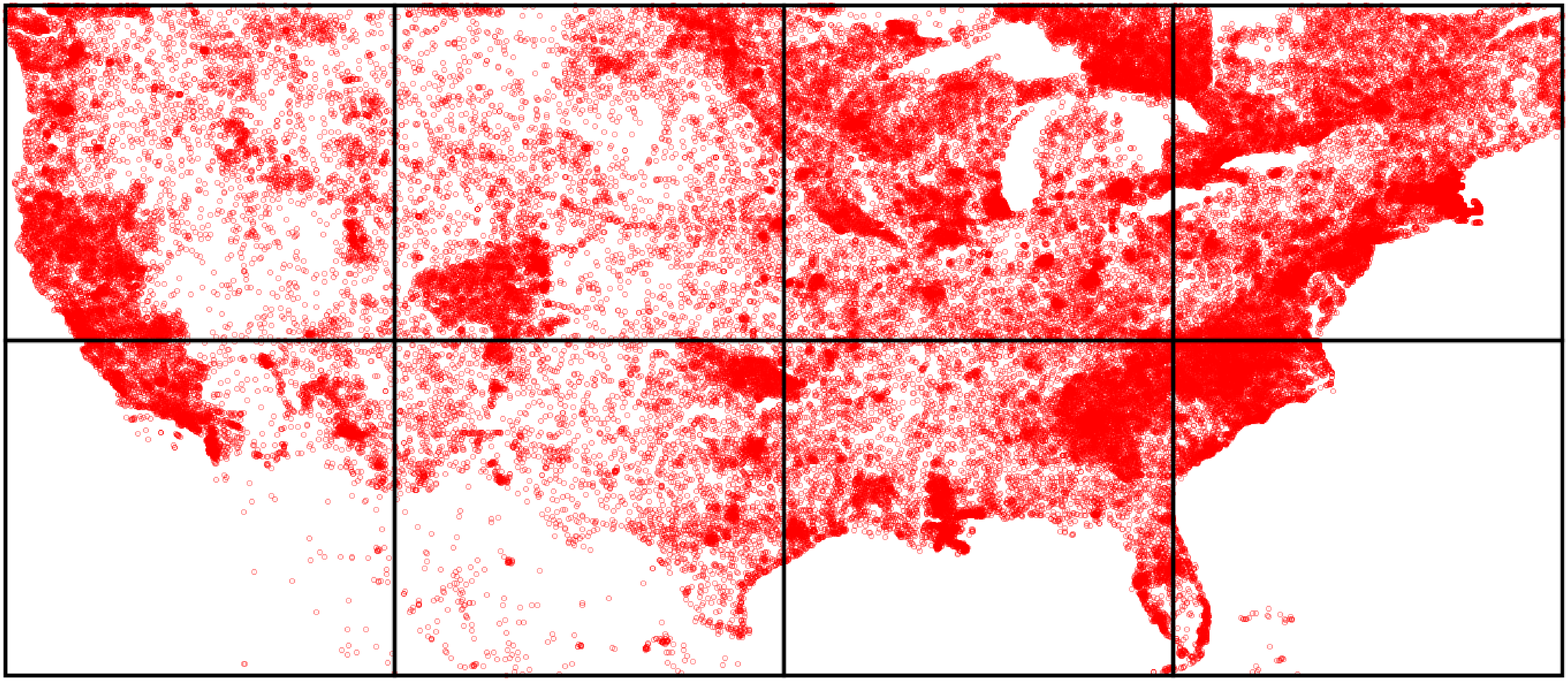}}%
	\subfigure[Quad-tree]{%
		\includegraphics[clip, width=0.3\hsize]{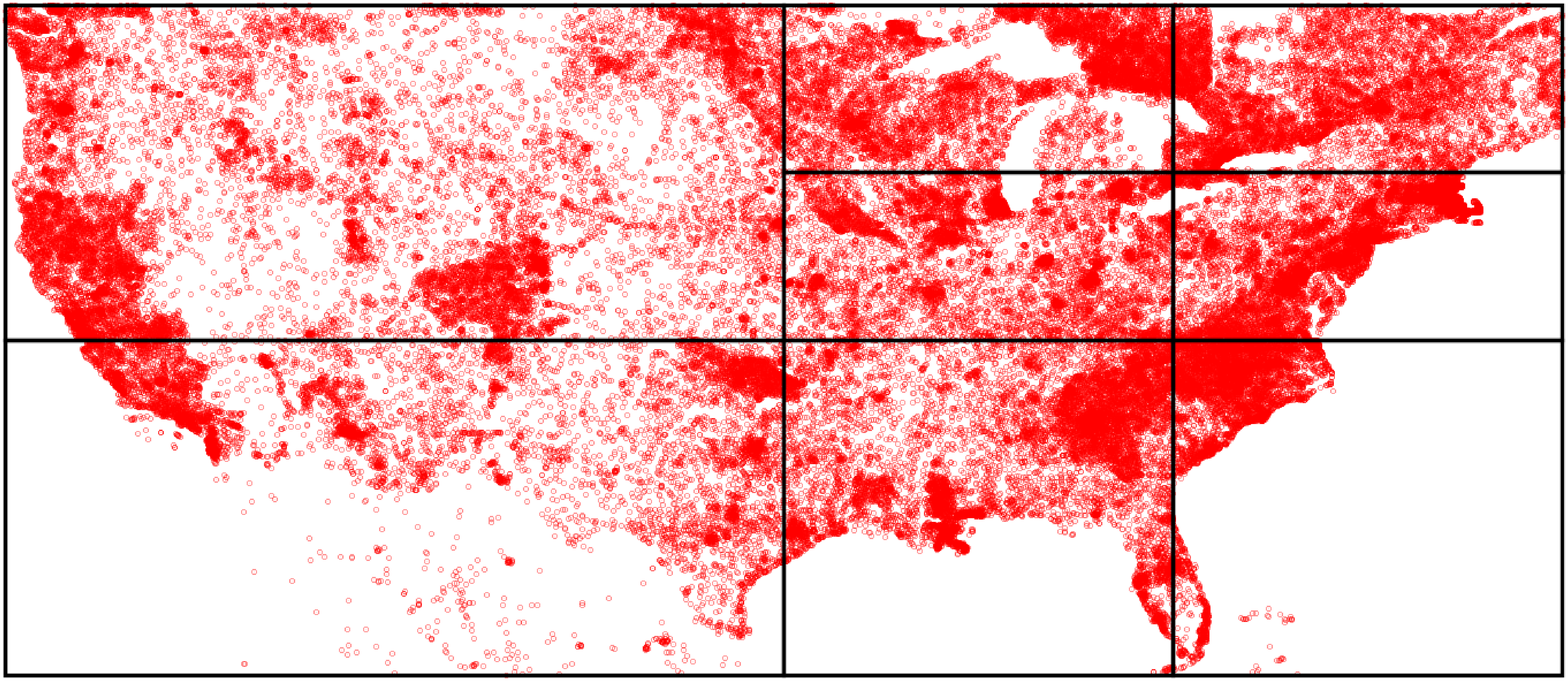}}%
	\\
	\subfigure[KDB-tree]{%
		\includegraphics[clip, width=0.3\hsize]{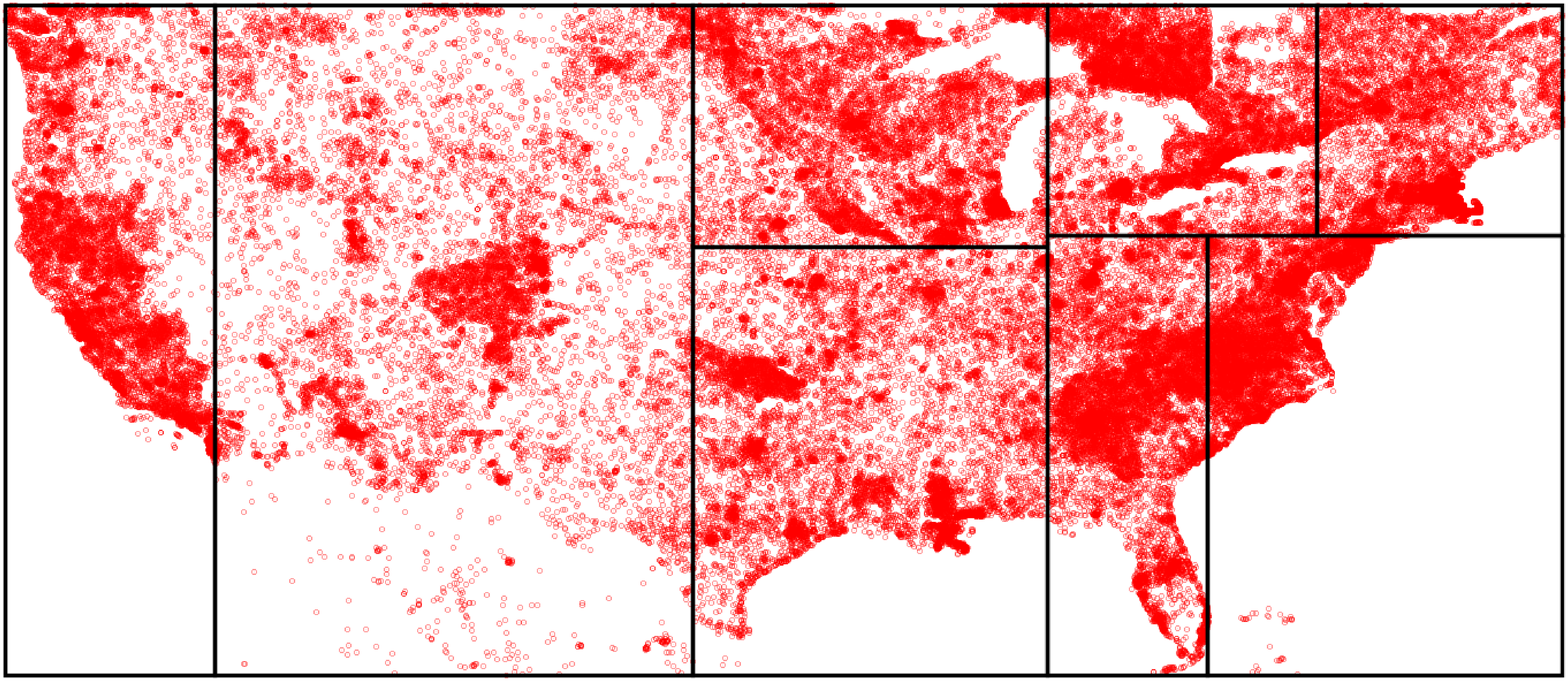}}%
	\subfigure[Demo]{%
		\includegraphics[clip, width=0.3\hsize]{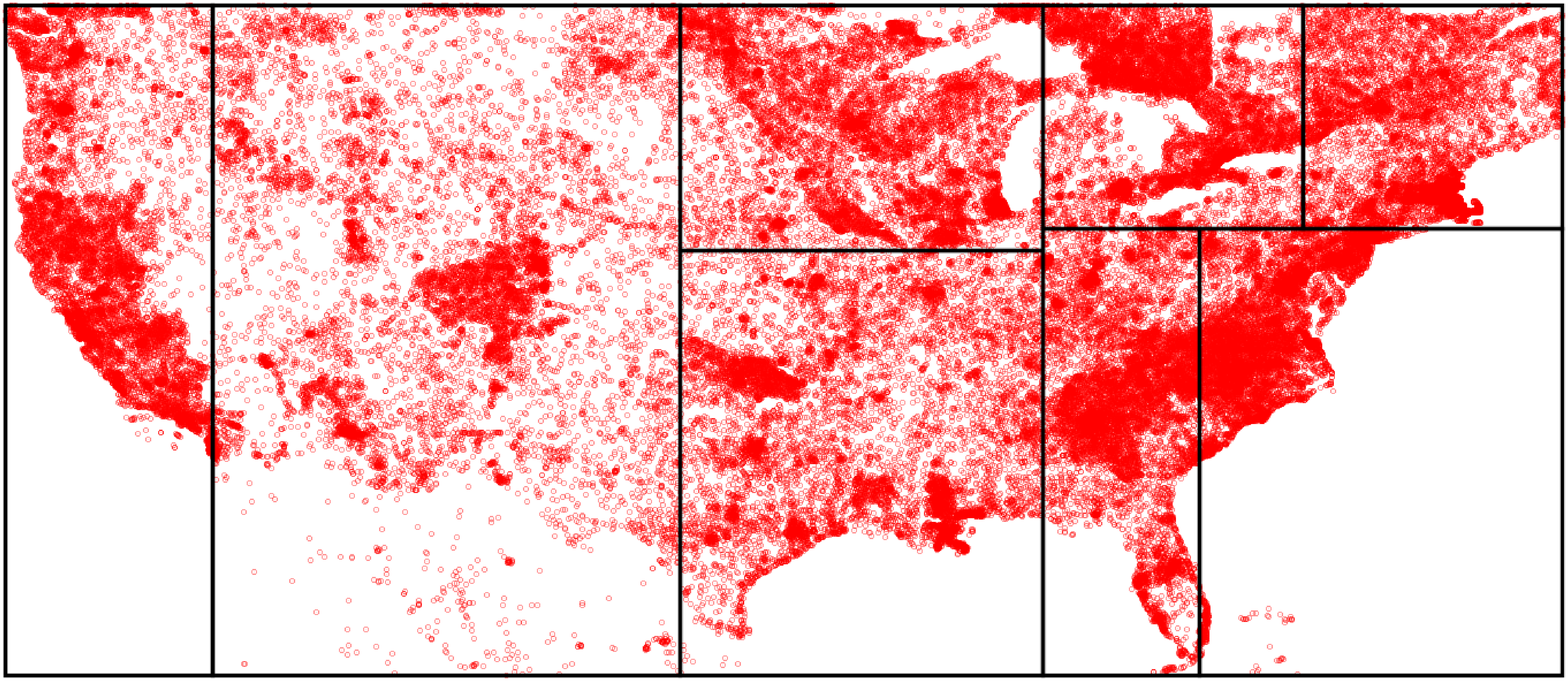}}%
	\subfigure[Ours]{%
		\includegraphics[clip, width=0.3\hsize]{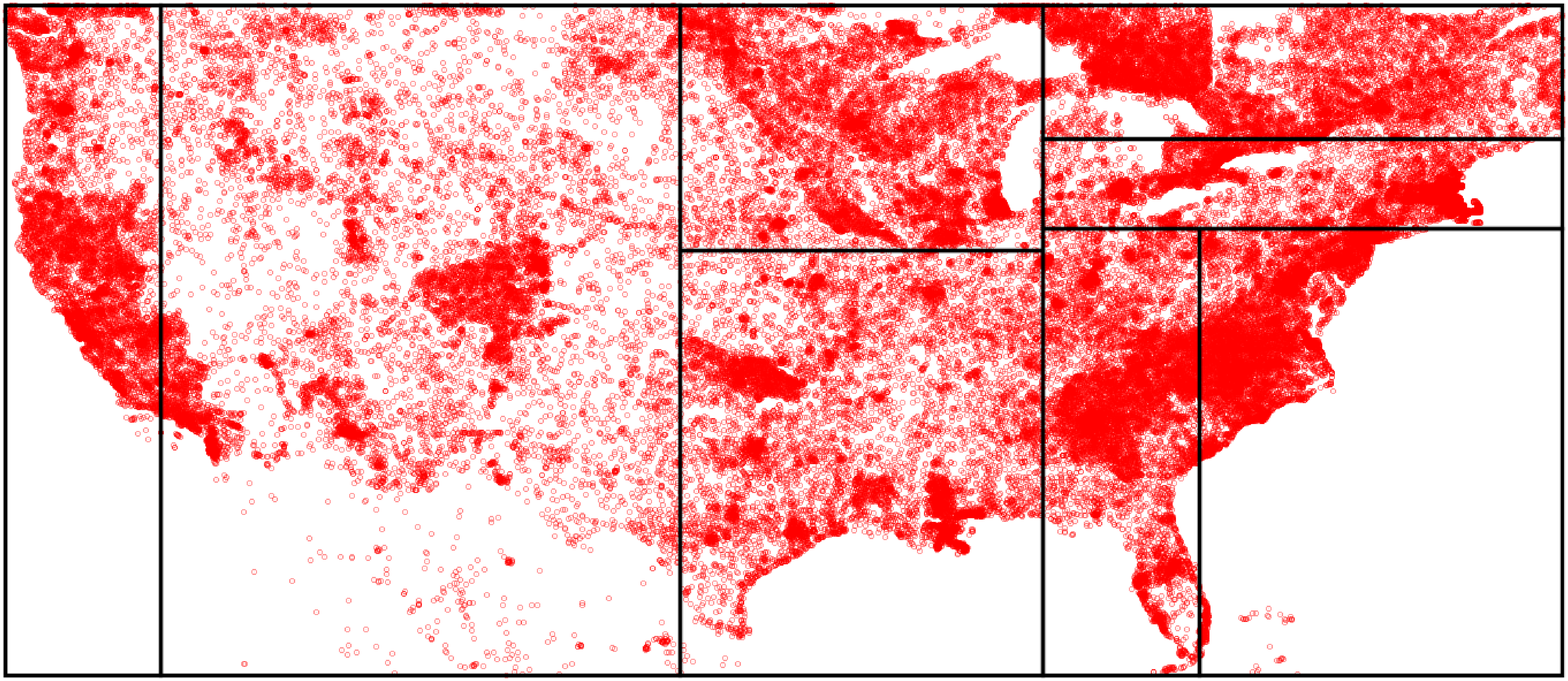}}%
	\caption{Comparison of partitions in OSM-US}
	\label{partitions_usa}
\end{figure*}

\smallskip
\noindent
{\bf Workload.}
Our workload was a set of distance joins, each of which finds all pairs of data within a given distance threshold.
Our workload included three distance thresholds, 500, 1000, and 5000 [meter] for US and SA, and 50, 100, and 500 [meters] for Imis, and we assumed two types of their frequencies; small skew and large skew.
The small skew has 25, 50, and 25 [\%] for the corresponding thresholds, respectively, while the large skew has 2, 3, and 95 [\%].

\smallskip
\noindent
{\bf Competitor.}
We used three spatial data partitioning methods; uniform grid, Quad-tree, and KDB-tree, which are originally implemented in Sedona.

\smallskip
\noindent
{\bf Hyper-parameter tuning.}
We divided the entire map into $30 \times 30$ grid cells.
In our $\epsilon$-greedy method, the selection probabilities of random and grid shift actions were $\epsilon_{r}=0.1$ and $\epsilon_{s}=0.2$, respectively.
We set the agent memory size to 10,000 and the demo memory size to 100. 
The mini-batch size for training was 32, and the percentage $\rho$ of demo data was $0.25$.
Furthermore, we set the number of steps for multi-step learning to 3 and the reward discount rate to $\gamma=0.99$.

The Q-network consists of two hidden layers that have 1200 and 600 nodes, respectively, from the input side. The input and output dimensions depend on the grid size.
In the case of our experimental setting, the input and output dimensions were 2833 (i.e., $31\times31\times3$) and 1800 (i.e., $30\times30\times2$), respectively.
As parameters during training, the learning rate $\eta$ was 0.001, and we used Adam as the optimizer.
The weights for the loss function were $\lambda_1=1.0$, $\lambda_2=1.0$, and $\lambda_3=10^{-5}$ according to the settings in the literature~\cite{DQfD}.

In the pruning execution of operators, when the run time of an operator was two times larger than that of $P_b$, we stopped running the workload and its reward was $0.2$.
We ran each operator in the given workload three times to mitigate the fluctuation of run time, and the median of these measurements was used for computing rewards.

\begin{table}[t]
\centering
\caption{Workload run time [sec]: The average run time of distance joins with each distance weighted by their frequencies. Bold and underline indicate the best and second best, respectively.}
\label{runtime-small-bias}
\centering
\begin{tabular}{l|lr|rrr|rr}\hline
Data & idx &skew  &Uniform &Quad &KDB &Demo &Ours \\\hline
\multirow{4}{*}{US}&\multirow{2}{*}{w/o}&\multirow{1}{*}{small}  &11.22 &8.931 &\underline{4.67} &4.90 &\textbf{4.56}  \\
&&\multirow{1}{*}{large}  &15.83 &13.29 &8.65 &8.95 &\textbf{6.71} \\\cline{2-8}
&\multirow{2}{*}{w/}&\multirow{1}{*}{small}  &3.35 &3.82 &\underline{3.00} &\underline{3.00} &\textbf{1.78}  \\
&&\multirow{1}{*}{large}  &7.93 &7.68 &7.32 &7.08 &\textbf{3.58}  \\\hline
\multirow{4}{*}{SA}&\multirow{2}{*}{w/o}&\multirow{1}{*}{small} &11.72 &11.91 &\underline{6.92} &7.22 &\textbf{6.90}\\
&&\multirow{1}{*}{large} &17.82 &18.21 &\textbf{13.21} &\underline{13.37} &13.60  \\\cline{2-8}
&\multirow{2}{*}{w/}&\multirow{1}{*}{small} &\underline{4.66} &5.21 &4.86 &5.09 &\textbf{3.52}  \\
&&\multirow{1}{*}{large} &11.29 &11.12 &11.19 &11.97 &\textbf{8.34} \\\hline
\multirow{4}{*}{Imis}&\multirow{2}{*}{w/o}&\multirow{1}{*}{small} &16.74 &12.69 &\textbf{8.77} &\underline{8.93} &9.49  \\
&&\multirow{1}{*}{large} &22.24 &19.81 &\underline{17.43} &17.49 &\textbf{15.57} \\\cline{2-8}
&\multirow{2}{*}{w/}&\multirow{1}{*}{small} &\underline{4.75} &5.03 &7.11 &7.12 &\textbf{3.60} \\
&&\multirow{1}{*}{large} &\underline{11.18} &11.84 &16.21 &15.87 &\textbf{7.84}  \\
\hline
\end{tabular}
\end{table}

\subsection{Experimental result}

\subsubsection{Example of partitions}
Figures \ref{partitions_usa} show the set of partitions constructed by each method on OSM-US.
Uniform grid and Quad-tree have a large bias in the number of data  in each partition because they ignore the number of data.
On the other hand, KDB-tree has a small bias, because it always divides the area so that each partition has (almost) the same number of data.
Demo was computed by KDB-tree, but there is a slight difference in the partitions because the boundaries are selected from the grid lines.
Ours is slightly different from Demo to achieve a better workload run time.
From this result, we can see that our algorithm imitates demo data and explores actions close to them. 
These examples show that our method can find effective partitions based on demo data.

\subsubsection{Effectiveness}
We evaluate the effectiveness of our method by the workload run time.
Table \ref{runtime-small-bias} shows the run time needed for running the workload. 
%distance joins with different thresholds and weighted run time by respective ratio. 
We evaluated three cases: small skew without local index, large skew without local index, and small skew with local index.
%We report the average run times in ten distance joins for each distance threshold.

Our method achieves the best performance among all the methods in all cases.
This indicates that our method can adaptively divide datasets into partitions according to workloads and computational environments. 
%Our method highly optimizes for 5000m in OSM-US and 500m in OSM-SA.
%We note that KDB-tree and Demo achieve better run time for several operators than ours because our algorithm aims at reducing the run time of the given workload instead of each operator. 
%The best partitions are different for each operator which results in poor performance at other distances.
When workloads have large skews, the difference between run times of ours and other methods becomes large, because our method optimizes partitions for workloads.
Our method reduced the workload run time up to 59.4\% compared with the second-best method. 

%Table 4 shows the run time when local indexes are applied to the experiments in Table 2. The local index is constructed on each machine for data retrieval. 

Compared to the difference between run times without and with local index, run times with local index are smaller than those without local index in all methods.
Our method achieves the smallest run time in both datasets, while KDB tree does not work well as uniform grid works in SA.
In US, our method reduces the run time by 41 \% compared to the second-best.
The reason why the local index largely benefits our method is that our method uses computational environments (i.e., including local index) to optimize partitions, while other methods do not. This result shows that spatial partitioning should be taken not only in data distribution but also in computational environments such as local index.
% Given a fixed number of partitions, a smaller number of partitions could achieve better performance.
% When the workers use local index, less than eight partitions are effective. 
% Our method can search for pseudo-optimal numbers of partitions but the existing methods do not change the partitions according to workloads and computational environments.  
%This is because our method can search for a pseudo-optimal number of partitions but the existing  
%Since the number of partitions is fixed, there are some partitions that appear unnecessary in Figure \ref{partitions-local-index}. This indicates that less than eight is actually suitable when local indexes are applied. In other words, the proposed method can search for a pseudo-optimal number of partitions.
We validate that DRL-based spatial data partitioning improves the performance of workloads and computational environments.

% Next, we evaluated the robustness of our method in two experiments: updating data and posing distance joins without the given workload.
% Fig. \ref{increase_datasize} shows the run time with increasing the data size from 5\% to 20\%.
% The existing methods and Demo construct partitions after increasing data, while our method applies partitions learned before increasing data (i.e., our method uses the same partitions for all data sizes).
% Our method achieves the best performance until data size increases 10\%. 
% When data size increases more than 10\%, the performance of our method is slightly lower than that of KDB-tree, but the difference is negligible.
% Table \ref{outsideworkloadruntime} shows the run time for distances outside the workload. 
% % Our method is effective enough for 4000m in OSM-US. 
% Our method can adapt to distances outside the workload as long as the distances are close to the given distances.

% We show the results for the distance joins without the given workload in our supplementary file.
% Briefly, our method achieves better performance than the existing methods in many cases even if we pose the distance thresholds without the given workload.

\begin{figure}[t]
  \centering
    \begin{tabular}{c}
      \begin{minipage}{0.7\linewidth}
        \centering
        \subfigure[Search efficiency]{
    	\includegraphics[width=1.0\linewidth]{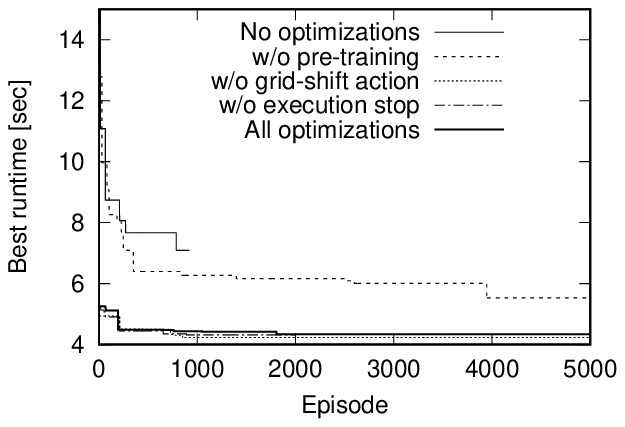}}%
    	%\label{search_efficiency}
      \end{minipage}\\
      \begin{minipage}{1.0\hsize}
        \centering
    	\subtable[Learning time]{
    \begin{tabular}{cc} \hline
        Methods & Learning time [hour] \\\hline
        No optimization  & 120 (Done 924 episodes) \\
        w/o pre-training  & 98.4 \\
        w/o grid-shift action & 72.5 \\
        w/o execution stop  & 120 (Done 2107 episodes) \\
        All optimization & 65.5 \\\hline
    \end{tabular}
    }
     \end{minipage}
    \end{tabular}
        \caption{Ablation study on the US dataset. We stopped training if a method did not finish 5,000 episodes within 120 hours.}
    \label{ablationstudy}
\end{figure} 

\subsubsection{Efficiency}
We evaluated the learning time and learning efficiency of our method.
For verifying the contribution of our learning strategy in Section \ref{strategy} to the learning performance, we conducted an ablation study using the following five patterns: 
no optimization, w/o pre-training, w/o grid-shift action, w/o execution stop, and all optimization.
%We use learning time and search efficiency as indices to show the relationship between workload run time and learning time.

Figure \ref{ablationstudy} shows the search efficiency and the learning time.
The search efficiency indicates the relationship between the number of episodes (x-axis) and the best run time during the training (y-axis); the shorter the run time during a small number of episodes, the more efficiently our method finds effective partitions. 

All optimization significantly outperforms no optimizations in both learning time and search efficiency.
This result indicates that our learning strategy is effective.
In w/o pre-training, the initial performance depends on random partitions, so it takes many episodes to find effective partitions.
We can confirm that the pre-training largely affects search efficiency.
In w/o grid-shift action, the search efficiency is similar to all optimization.
However, it often constructs less-effective partitions due to random actions, though the searches by random action have the potential to find better actions.
This causes increasing the training time because less-effective partitions often take a large run time.
In w/o execution stop, the learning time significantly increases. 
By comparing all optimization with w/o execution stop, we can confirm that our method sufficiently trains our models without computing strict rewards.
%by comparing all optimization with w/o execution stop.

These results show that our learning strategy significantly contributes to efficient learning in our problem formulation.

\section{Conclusion and Open Challenges}
\label{conclusion}
We studied deep reinforcement learning for spatial partitioning.
We first formalized the spatial partitioning problem as deep reinforcement learning, and we developed a learning algorithm for the problem.
Our algorithm combines Deep Q-network with imitation learning to explore actions effectively and efficiently.
It includes three optimization techniques: pre-training by effective demo data partitioning, effective new action choice, and pruning run time measurement.
Through our experiment using Apache Sedona and two real datasets, we validated that (i) our method is effective and accelerates spatial join processing and (ii) our optimization techniques reduce its learning time.

There are many open challenges to extending our work because we first studied learned spatial data partitioning. 
Promising challenges are to reduce learning time, select the optimal number of partitions instead of using the number of computers, and develop further optimal learning frameworks.
Furthermore, it is interesting to investigate the performance of other distributed parallel processing systems and spatial queries.

%Furthermore, we simplify the number of partitions to the number of machines, so we plan to learn the number of partitions dynamically.

\section*{Acknowledgement}
This work was supported by JSPS KAKENHI Grant Numbers JP20H00584 and JP22H03700.

\bibliographystyle{ACM-Reference-Format}
\bibliography{bibliography}

%%% -*-BibTeX-*-
%%% Do NOT edit. File created by BibTeX with style
%%% ACM-Reference-Format-Journals [18-Jan-2012].

\begin{thebibliography}{37}

%%% ====================================================================
%%% NOTE TO THE USER: you can override these defaults by providing
%%% customized versions of any of these macros before the \bibliography
%%% command.  Each of them MUST provide its own final punctuation,
%%% except for \shownote{}, \showDOI{}, and \showURL{}.  The latter two
%%% do not use final punctuation, in order to avoid confusing it with
%%% the Web address.
%%%
%%% To suppress output of a particular field, define its macro to expand
%%% to an empty string, or better, \unskip, like this:
%%%
%%% \newcommand{\showDOI}[1]{\unskip}   % LaTeX syntax
%%%
%%% \def \showDOI #1{\unskip}           % plain TeX syntax
%%%
%%% ====================================================================

\ifx \showCODEN    \undefined \def \showCODEN     #1{\unskip}     \fi
\ifx \showDOI      \undefined \def \showDOI       #1{#1}\fi
\ifx \showISBNx    \undefined \def \showISBNx     #1{\unskip}     \fi
\ifx \showISBNxiii \undefined \def \showISBNxiii  #1{\unskip}     \fi
\ifx \showISSN     \undefined \def \showISSN      #1{\unskip}     \fi
\ifx \showLCCN     \undefined \def \showLCCN      #1{\unskip}     \fi
\ifx \shownote     \undefined \def \shownote      #1{#1}          \fi
\ifx \showarticletitle \undefined \def \showarticletitle #1{#1}   \fi
\ifx \showURL      \undefined \def \showURL       {\relax}        \fi
% The following commands are used for tagged output and should be
% invisible to TeX
\providecommand\bibfield[2]{#2}
\providecommand\bibinfo[2]{#2}
\providecommand\natexlab[1]{#1}
\providecommand\showeprint[2][]{arXiv:#2}

\bibitem[Aji et~al\mbox{.}(2013)]%
        {aji2013hadoop}
\bibfield{author}{\bibinfo{person}{Ablimit Aji}, \bibinfo{person}{Fusheng
  Wang}, \bibinfo{person}{Hoang Vo}, \bibinfo{person}{Rubao Lee},
  \bibinfo{person}{Qiaoling Liu}, \bibinfo{person}{Xiaodong Zhang}, {and}
  \bibinfo{person}{Joel Saltz}.} \bibinfo{year}{2013}\natexlab{}.
\newblock \showarticletitle{{Hadoop-GIS}: a high performance spatial data
  warehousing system over mapreduce}.
\newblock \bibinfo{journal}{\emph{PVLDB}} \bibinfo{volume}{6},
  \bibinfo{number}{11} (\bibinfo{year}{2013}), \bibinfo{pages}{1009--1020}.
\newblock


\bibitem[Aly et~al\mbox{.}(2015)]%
        {aly2015aqwa}
\bibfield{author}{\bibinfo{person}{Ahmed~M Aly}, \bibinfo{person}{Ahmed~R
  Mahmood}, \bibinfo{person}{Mohamed~S Hassan}, \bibinfo{person}{Walid~G Aref},
  \bibinfo{person}{Mourad Ouzzani}, \bibinfo{person}{Hazem Elmeleegy}, {and}
  \bibinfo{person}{Thamir Qadah}.} \bibinfo{year}{2015}\natexlab{}.
\newblock \showarticletitle{Aqwa: adaptive query workload aware partitioning of
  big spatial data}.
\newblock \bibinfo{journal}{\emph{PVLDB}} \bibinfo{volume}{8},
  \bibinfo{number}{13} (\bibinfo{year}{2015}), \bibinfo{pages}{2062--2073}.
\newblock


\bibitem[{Apache Sedona}(2021)]%
        {sedona}
\bibfield{author}{\bibinfo{person}{{Apache Sedona}}.}
  \bibinfo{year}{2021}\natexlab{}.
\newblock \bibinfo{howpublished}{https://sedona.apache.org/}.
\newblock


\bibitem[Asis et~al\mbox{.}(2018)]%
        {deasis2018multistep}
\bibfield{author}{\bibinfo{person}{Kristopher~De Asis},
  \bibinfo{person}{J.~Fernando Hernandez-Garcia}, \bibinfo{person}{G.~Zacharias
  Holland}, {and} \bibinfo{person}{Richard~S. Sutton}.}
  \bibinfo{year}{2018}\natexlab{}.
\newblock \showarticletitle{Multi-step Reinforcement Learning: A Unifying
  Algorithm}.
\newblock \bibinfo{journal}{\emph{arXiv}} (\bibinfo{year}{2018}).
\newblock


\bibitem[Ding et~al\mbox{.}(2020)]%
        {ding2020tsunami}
\bibfield{author}{\bibinfo{person}{Jialin Ding}, \bibinfo{person}{Vikram
  Nathan}, \bibinfo{person}{Mohammad Alizadeh}, {and} \bibinfo{person}{Tim
  Kraska}.} \bibinfo{year}{2020}\natexlab{}.
\newblock \showarticletitle{Tsunami: A learned multi-dimensional index for
  correlated data and skewed workloads}.
\newblock \bibinfo{journal}{\emph{PVLDB}} (\bibinfo{year}{2020}),
  \bibinfo{pages}{74--86}.
\newblock


\bibitem[Eldawy et~al\mbox{.}(2015)]%
        {eldawy2015spatial}
\bibfield{author}{\bibinfo{person}{Ahmed Eldawy}, \bibinfo{person}{Louai
  Alarabi}, {and} \bibinfo{person}{Mohamed~F Mokbel}.}
  \bibinfo{year}{2015}\natexlab{}.
\newblock \showarticletitle{Spatial partitioning techniques in SpatialHadoop}.
\newblock \bibinfo{journal}{\emph{PVLDB}} \bibinfo{volume}{8},
  \bibinfo{number}{12} (\bibinfo{year}{2015}), \bibinfo{pages}{1602--1605}.
\newblock


\bibitem[Eldawy and Mokbel(2015)]%
        {spatialhadoop}
\bibfield{author}{\bibinfo{person}{Ahmed Eldawy} {and}
  \bibinfo{person}{Mohamed~F. Mokbel}.} \bibinfo{year}{2015}\natexlab{}.
\newblock \showarticletitle{{SpatialHadoop}: A {MapReduce} framework for
  spatial data}. In \bibinfo{booktitle}{\emph{ICDE}}.
  \bibinfo{pages}{1352--1363}.
\newblock


\bibitem[Finkel and Bentley(1974)]%
        {quad-tree}
\bibfield{author}{\bibinfo{person}{R.~A. Finkel} {and} \bibinfo{person}{J.~L.
  Bentley}.} \bibinfo{year}{1974}\natexlab{}.
\newblock \showarticletitle{Quad Trees a Data Structure for Retrieval on
  Composite Keys}.
\newblock \bibinfo{journal}{\emph{The Acta Informatica}} \bibinfo{volume}{4},
  \bibinfo{number}{1} (\bibinfo{year}{1974}), \bibinfo{pages}{1--9}.
\newblock


\bibitem[Gu et~al\mbox{.}(2021)]%
        {https://doi.org/10.48550/arxiv.2103.04541}
\bibfield{author}{\bibinfo{person}{Tu Gu}, \bibinfo{person}{Kaiyu Feng},
  \bibinfo{person}{Gao Cong}, \bibinfo{person}{Cheng Long},
  \bibinfo{person}{Zheng Wang}, {and} \bibinfo{person}{Sheng Wang}.}
  \bibinfo{year}{2021}\natexlab{}.
\newblock \bibinfo{title}{A Reinforcement Learning Based R-Tree for Spatial
  Data Indexing in Dynamic Environments}.
\newblock
\newblock


\bibitem[Hester et~al\mbox{.}(2018)]%
        {DQfD}
\bibfield{author}{\bibinfo{person}{Todd Hester}, \bibinfo{person}{Matej
  Vecerik}, \bibinfo{person}{Olivier Pietquin}, \bibinfo{person}{Marc Lanctot},
  \bibinfo{person}{Tom Schaul}, \bibinfo{person}{Bilal Piot},
  \bibinfo{person}{Dan Horgan}, \bibinfo{person}{John Quan},
  \bibinfo{person}{Andrew Sendonaris}, \bibinfo{person}{Ian Osband},
  \bibinfo{person}{Gabriel Dulac-Arnold}, \bibinfo{person}{John Agapiou},
  \bibinfo{person}{Joel Leibo}, {and} \bibinfo{person}{Audrunas Gruslys}.}
  \bibinfo{year}{2018}\natexlab{}.
\newblock \showarticletitle{Deep Q-learning From Demonstrations}. In
  \bibinfo{booktitle}{\emph{AAAI}}. \bibinfo{pages}{3223--3230}.
\newblock


\bibitem[Hilprecht et~al\mbox{.}(2020)]%
        {10.1145/3318464.3389704}
\bibfield{author}{\bibinfo{person}{Benjamin Hilprecht},
  \bibinfo{person}{Carsten Binnig}, {and} \bibinfo{person}{Uwe R\"{o}hm}.}
  \bibinfo{year}{2020}\natexlab{}.
\newblock \showarticletitle{Learning a Partitioning Advisor for Cloud
  Databases}. In \bibinfo{booktitle}{\emph{SIGMOD}}. \bibinfo{pages}{143--157}.
\newblock


\bibitem[Julian and Kochenderfer(2019)]%
        {julian2019distributed}
\bibfield{author}{\bibinfo{person}{Kyle~D Julian} {and}
  \bibinfo{person}{Mykel~J Kochenderfer}.} \bibinfo{year}{2019}\natexlab{}.
\newblock \showarticletitle{Distributed wildfire surveillance with autonomous
  aircraft using deep reinforcement learning}.
\newblock \bibinfo{journal}{\emph{Journal of Guidance, Control, and Dynamics}}
  \bibinfo{volume}{42}, \bibinfo{number}{8} (\bibinfo{year}{2019}),
  \bibinfo{pages}{1768--1778}.
\newblock


\bibitem[Kaelbling et~al\mbox{.}(1996)]%
        {reinforcement_learning}
\bibfield{author}{\bibinfo{person}{Leslie~Pack Kaelbling},
  \bibinfo{person}{Michael~L Littman}, {and} \bibinfo{person}{Andrew~W Moore}.}
  \bibinfo{year}{1996}\natexlab{}.
\newblock \showarticletitle{Reinforcement learning: A survey}.
\newblock \bibinfo{journal}{\emph{JAIR}}  \bibinfo{volume}{4}
  (\bibinfo{year}{1996}), \bibinfo{pages}{237--285}.
\newblock


\bibitem[Lan et~al\mbox{.}(2020)]%
        {10.1145/3340531.3412106}
\bibfield{author}{\bibinfo{person}{Hai Lan}, \bibinfo{person}{Zhifeng Bao},
  {and} \bibinfo{person}{Yuwei Peng}.} \bibinfo{year}{2020}\natexlab{}.
\newblock \showarticletitle{An Index Advisor Using Deep Reinforcement
  Learning}. In \bibinfo{booktitle}{\emph{CIKM}}. \bibinfo{pages}{2105--2108}.
\newblock


\bibitem[Li et~al\mbox{.}(2020)]%
        {li2020lisa}
\bibfield{author}{\bibinfo{person}{Pengfei Li}, \bibinfo{person}{Hua Lu},
  \bibinfo{person}{Qian Zheng}, \bibinfo{person}{Long Yang}, {and}
  \bibinfo{person}{Gang Pan}.} \bibinfo{year}{2020}\natexlab{}.
\newblock \showarticletitle{LISA: A learned index structure for spatial data}.
  In \bibinfo{booktitle}{\emph{SIGMOD}}. \bibinfo{pages}{2119--2133}.
\newblock


\bibitem[Li(2017)]%
        {li2017deep}
\bibfield{author}{\bibinfo{person}{Yuxi Li}.} \bibinfo{year}{2017}\natexlab{}.
\newblock \showarticletitle{Deep reinforcement learning: An overview}.
\newblock \bibinfo{journal}{\emph{arXiv}} (\bibinfo{year}{2017}).
\newblock


\bibitem[Liu et~al\mbox{.}(2019)]%
        {liu2019geo}
\bibfield{author}{\bibinfo{person}{Wei Liu}, \bibinfo{person}{Zhi-Jie Wang},
  \bibinfo{person}{Bin Yao}, {and} \bibinfo{person}{Jian Yin}.}
  \bibinfo{year}{2019}\natexlab{}.
\newblock \showarticletitle{Geo-ALM: POI Recommendation by Fusing Geographical
  Information and Adversarial Learning Mechanism.}. In
  \bibinfo{booktitle}{\emph{IJCAI}}, Vol.~\bibinfo{volume}{7}.
  \bibinfo{pages}{1807--1813}.
\newblock


\bibitem[Mnih et~al\mbox{.}(2015)]%
        {DQN}
\bibfield{author}{\bibinfo{person}{Volodymyr Mnih}, \bibinfo{person}{Koray
  Kavukcuoglu}, \bibinfo{person}{David Silver}, \bibinfo{person}{Andrei~A.
  Rusu}, \bibinfo{person}{Joel Veness}, \bibinfo{person}{Marc~G. Bellemare},
  \bibinfo{person}{Alex Graves}, \bibinfo{person}{Martin Riedmiller},
  \bibinfo{person}{Andreas~K. Fidjeland}, \bibinfo{person}{Georg Ostrovski},
  \bibinfo{person}{Stig Petersen}, \bibinfo{person}{Charles Beattie},
  \bibinfo{person}{Amir Sadik}, \bibinfo{person}{Ioannis Antonoglou},
  \bibinfo{person}{Helen King}, \bibinfo{person}{Dharshan Kumaran},
  \bibinfo{person}{Daan Wierstra}, \bibinfo{person}{Shane Legg}, {and}
  \bibinfo{person}{Demis Hassabis}.} \bibinfo{year}{2015}\natexlab{}.
\newblock \showarticletitle{Human-level control through deep reinforcement
  learning}.
\newblock \bibinfo{journal}{\emph{Nature}} \bibinfo{volume}{518},
  \bibinfo{number}{7540} (\bibinfo{year}{2015}), \bibinfo{pages}{529--533}.
\newblock


\bibitem[Nair et~al\mbox{.}(2015)]%
        {gorilla}
\bibfield{author}{\bibinfo{person}{Arun Nair}, \bibinfo{person}{Praveen
  Srinivasan}, \bibinfo{person}{Sam Blackwell}, \bibinfo{person}{Cagdas
  Alcicek}, \bibinfo{person}{Rory Fearon}, \bibinfo{person}{Alessandro~De
  Maria}, \bibinfo{person}{Vedavyas Panneershelvam}, \bibinfo{person}{Mustafa
  Suleyman}, \bibinfo{person}{Charles Beattie}, \bibinfo{person}{Stig
  Petersen}, \bibinfo{person}{Shane Legg}, \bibinfo{person}{Volodymyr Mnih},
  \bibinfo{person}{Koray Kavukcuoglu}, {and} \bibinfo{person}{David Silver}.}
  \bibinfo{year}{2015}\natexlab{}.
\newblock \showarticletitle{Massively Parallel Methods for Deep Reinforcement
  Learning}.
\newblock \bibinfo{journal}{\emph{arXiv}} (\bibinfo{year}{2015}).
\newblock


\bibitem[Nathan et~al\mbox{.}(2020)]%
        {nathan2020learning}
\bibfield{author}{\bibinfo{person}{Vikram Nathan}, \bibinfo{person}{Jialin
  Ding}, \bibinfo{person}{Mohammad Alizadeh}, {and} \bibinfo{person}{Tim
  Kraska}.} \bibinfo{year}{2020}\natexlab{}.
\newblock \showarticletitle{Learning multi-dimensional indexes}. In
  \bibinfo{booktitle}{\emph{SIGMOD}}. \bibinfo{pages}{985--1000}.
\newblock


\bibitem[Pan et~al\mbox{.}(2019)]%
        {pan2019deep}
\bibfield{author}{\bibinfo{person}{Ling Pan}, \bibinfo{person}{Qingpeng Cai},
  \bibinfo{person}{Zhixuan Fang}, \bibinfo{person}{Pingzhong Tang}, {and}
  \bibinfo{person}{Longbo Huang}.} \bibinfo{year}{2019}\natexlab{}.
\newblock \showarticletitle{A deep reinforcement learning framework for
  rebalancing dockless bike sharing systems}. In
  \bibinfo{booktitle}{\emph{AAAI}}, Vol.~\bibinfo{volume}{33}.
  \bibinfo{pages}{1393--1400}.
\newblock


\bibitem[Pang et~al\mbox{.}(2020)]%
        {pang2020intercity}
\bibfield{author}{\bibinfo{person}{Yanbo Pang}, \bibinfo{person}{Kota
  Tsubouchi}, \bibinfo{person}{Takahiro Yabe}, {and} \bibinfo{person}{Yoshihide
  Sekimoto}.} \bibinfo{year}{2020}\natexlab{}.
\newblock \showarticletitle{Intercity Simulation of Human Mobility at Rare
  Events via Reinforcement Learning}. In
  \bibinfo{booktitle}{\emph{SIGSPATIAL}}. \bibinfo{pages}{293--302}.
\newblock


\bibitem[Qi et~al\mbox{.}(2020)]%
        {qi2020effectively}
\bibfield{author}{\bibinfo{person}{Jianzhong Qi}, \bibinfo{person}{Guanli Liu},
  \bibinfo{person}{Christian~S Jensen}, {and} \bibinfo{person}{Lars Kulik}.}
  \bibinfo{year}{2020}\natexlab{}.
\newblock \showarticletitle{Effectively learning spatial indices}.
\newblock \bibinfo{journal}{\emph{PVLDB}} (\bibinfo{year}{2020}),
  \bibinfo{pages}{2341--2354}.
\newblock


\bibitem[Robinson(1981)]%
        {kdb-tree}
\bibfield{author}{\bibinfo{person}{John~T. Robinson}.}
  \bibinfo{year}{1981}\natexlab{}.
\newblock \showarticletitle{The {K-D-B-Tree}: A Search Structure for Large
  Multidimensional Dynamic Indexes}. In \bibinfo{booktitle}{\emph{SIGMOD}}.
  \bibinfo{pages}{10--18}.
\newblock


\bibitem[Sasaki(2021)]%
        {sasaki2021survey}
\bibfield{author}{\bibinfo{person}{Yuya Sasaki}.}
  \bibinfo{year}{2021}\natexlab{}.
\newblock \showarticletitle{A Survey on IoT Big Data Analytic Systems: Current
  and Future}.
\newblock \bibinfo{journal}{\emph{IEEE Internet of Things Journal}}
  (\bibinfo{year}{2021}).
\newblock


\bibitem[Sasaki et~al\mbox{.}(2018)]%
        {sasaki2020sequenced}
\bibfield{author}{\bibinfo{person}{Yuya Sasaki}, \bibinfo{person}{Yoshiharu
  Ishikawa}, \bibinfo{person}{Yasuhiro Fujiwara}, {and} \bibinfo{person}{Makoto
  Onizuka}.} \bibinfo{year}{2018}\natexlab{}.
\newblock \showarticletitle{Sequenced route query with semantic hierarchy}. In
  \bibinfo{booktitle}{\emph{EDBT}}. \bibinfo{pages}{37--48}.
\newblock


\bibitem[Schaul et~al\mbox{.}(2016)]%
        {PrioritizedExperienceReplay}
\bibfield{author}{\bibinfo{person}{Tom Schaul}, \bibinfo{person}{John Quan},
  \bibinfo{person}{Ioannis Antonoglou}, {and} \bibinfo{person}{David Silver}.}
  \bibinfo{year}{2016}\natexlab{}.
\newblock \showarticletitle{Prioritized Experience Replay}.
\newblock \bibinfo{journal}{\emph{arXiv}} (\bibinfo{year}{2016}).
\newblock
\showeprint{1511.05952}


\bibitem[Sevim et~al\mbox{.}(2021)]%
        {sevim2021brief}
\bibfield{author}{\bibinfo{person}{Akil Sevim},
  \bibinfo{person}{Mehnaz~Tabassum Mahin}, \bibinfo{person}{Tin Vu},
  \bibinfo{person}{Ian Maxon}, \bibinfo{person}{Ahmed Eldawy},
  \bibinfo{person}{Michael Carey}, {and} \bibinfo{person}{Vassilis Tsotras}.}
  \bibinfo{year}{2021}\natexlab{}.
\newblock \showarticletitle{A brief introduction to geospatial big data
  analytics with apache AsterixDB}. In \bibinfo{booktitle}{\emph{International
  Workshop on APIs and Libraries for Geospatial Data Science}}.
  \bibinfo{pages}{1--2}.
\newblock


\bibitem[Tang et~al\mbox{.}(2016)]%
        {tang2016locationspark}
\bibfield{author}{\bibinfo{person}{Mingjie Tang}, \bibinfo{person}{Yongyang
  Yu}, \bibinfo{person}{Qutaibah~M Malluhi}, \bibinfo{person}{Mourad Ouzzani},
  {and} \bibinfo{person}{Walid~G Aref}.} \bibinfo{year}{2016}\natexlab{}.
\newblock \showarticletitle{Locationspark: A distributed in-memory data
  management system for big spatial data}.
\newblock \bibinfo{journal}{\emph{PVLDB}} \bibinfo{volume}{9},
  \bibinfo{number}{13} (\bibinfo{year}{2016}), \bibinfo{pages}{1565--1568}.
\newblock


\bibitem[Thrun and Littman(2000)]%
        {thrun2000reinforcement}
\bibfield{author}{\bibinfo{person}{Sebastian Thrun} {and}
  \bibinfo{person}{Michael~L Littman}.} \bibinfo{year}{2000}\natexlab{}.
\newblock \showarticletitle{Reinforcement learning: an introduction}.
\newblock \bibinfo{journal}{\emph{AI Magazine}} \bibinfo{volume}{21},
  \bibinfo{number}{1} (\bibinfo{year}{2000}), \bibinfo{pages}{103--103}.
\newblock


\bibitem[Vu et~al\mbox{.}(2021a)]%
        {vu2021learned}
\bibfield{author}{\bibinfo{person}{Tin Vu}, \bibinfo{person}{Alberto Belussi},
  \bibinfo{person}{Sara Migliorini}, {and} \bibinfo{person}{Ahmed Eldawy}.}
  \bibinfo{year}{2021}\natexlab{a}.
\newblock \showarticletitle{A Learned Query Optimizer for Spatial Join}. In
  \bibinfo{booktitle}{\emph{SIGSPATIAL}}. \bibinfo{pages}{458--467}.
\newblock


\bibitem[Vu et~al\mbox{.}(2020)]%
        {10.1145/3402126}
\bibfield{author}{\bibinfo{person}{Tin Vu}, \bibinfo{person}{Alberto Belussi},
  \bibinfo{person}{Sara Migliorini}, {and} \bibinfo{person}{Ahmed Eldway}.}
  \bibinfo{year}{2020}\natexlab{}.
\newblock \showarticletitle{Using Deep Learning for Big Spatial Data
  Partitioning}.
\newblock \bibinfo{journal}{\emph{TSAS}}  \bibinfo{volume}{7}
  (\bibinfo{date}{08} \bibinfo{year}{2020}), \bibinfo{pages}{1--37}.
\newblock


\bibitem[Vu et~al\mbox{.}(2021b)]%
        {vu2021incremental}
\bibfield{author}{\bibinfo{person}{Tin Vu}, \bibinfo{person}{Ahmed Eldawy},
  \bibinfo{person}{Vagelis Hristidis}, {and} \bibinfo{person}{Vassilis
  Tsotras}.} \bibinfo{year}{2021}\natexlab{b}.
\newblock \showarticletitle{Incremental partitioning for efficient spatial data
  analytics}.
\newblock \bibinfo{journal}{\emph{PVLDB}} \bibinfo{volume}{15},
  \bibinfo{number}{3} (\bibinfo{year}{2021}), \bibinfo{pages}{713--726}.
\newblock


\bibitem[Watkins and Dayan(2004)]%
        {q-learning}
\bibfield{author}{\bibinfo{person}{Christopher~JCH Watkins} {and}
  \bibinfo{person}{Peter Dayan}.} \bibinfo{year}{2004}\natexlab{}.
\newblock \showarticletitle{Technical Note: Q-Learning}.
\newblock \bibinfo{journal}{\emph{The Machine Learning}}  \bibinfo{volume}{8}
  (\bibinfo{year}{2004}), \bibinfo{pages}{279--292}.
\newblock


\bibitem[Wei et~al\mbox{.}(2018)]%
        {wei2018intellilight}
\bibfield{author}{\bibinfo{person}{Hua Wei}, \bibinfo{person}{Guanjie Zheng},
  \bibinfo{person}{Huaxiu Yao}, {and} \bibinfo{person}{Zhenhui Li}.}
  \bibinfo{year}{2018}\natexlab{}.
\newblock \showarticletitle{Intellilight: A reinforcement learning approach for
  intelligent traffic light control}. In \bibinfo{booktitle}{\emph{SIGKDD}}.
  \bibinfo{pages}{2496--2505}.
\newblock


\bibitem[Xie et~al\mbox{.}(2016)]%
        {xie2016simba}
\bibfield{author}{\bibinfo{person}{Dong Xie}, \bibinfo{person}{Feifei Li},
  \bibinfo{person}{Bin Yao}, \bibinfo{person}{Gefei Li}, \bibinfo{person}{Liang
  Zhou}, {and} \bibinfo{person}{Minyi Guo}.} \bibinfo{year}{2016}\natexlab{}.
\newblock \showarticletitle{Simba: Efficient in-memory spatial analytics}. In
  \bibinfo{booktitle}{\emph{Proceedings of the SIGMOD}}.
  \bibinfo{pages}{1071--1085}.
\newblock


\bibitem[Yang et~al\mbox{.}(2020)]%
        {yang2020qd}
\bibfield{author}{\bibinfo{person}{Zongheng Yang}, \bibinfo{person}{Badrish
  Chandramouli}, \bibinfo{person}{Chi Wang}, \bibinfo{person}{Johannes Gehrke},
  \bibinfo{person}{Yinan Li}, \bibinfo{person}{Umar~Farooq Minhas},
  \bibinfo{person}{Per-{\AA}ke Larson}, \bibinfo{person}{Donald Kossmann},
  {and} \bibinfo{person}{Rajeev Acharya}.} \bibinfo{year}{2020}\natexlab{}.
\newblock \showarticletitle{Qd-tree: Learning data layouts for big data
  analytics}. In \bibinfo{booktitle}{\emph{SIGMOD}}. \bibinfo{pages}{193--208}.
\newblock


\end{thebibliography}

\end{document}